\documentclass[10pt,journal,compsoc]{IEEEtran}
	\usepackage{cite}
	\ifCLASSINFOpdf
	\else
	\fi
	\usepackage{tabularx}
\usepackage{booktabs}
\usepackage{longtable}
\usepackage{multirow}
\usepackage{colortbl}
\usepackage{caption}
\usepackage{tabu}%
\usepackage{longtable}
\usepackage{hhline}
\usepackage{tabularx,colortbl}
\usepackage{mathtools}
\usepackage{eqnarray,amsmath}
\usepackage{amsmath}
\usepackage{hyperref} % for autoref
\usepackage{cite}
\usepackage{amsmath,amssymb,amsfonts}
\usepackage{algorithmic}
\usepackage{graphicx}
\usepackage{enumitem}
\usepackage[utf8]{inputenc}
 \usepackage{subfig}
\usepackage{graphicx}
\usepackage{comment}
\usepackage{makecell}
\usepackage{wasysym}
\usepackage{mathrsfs}
\usepackage{amsthm}
\usepackage{bm}

\usepackage{soul}
\usepackage[export]{adjustbox}

\usepackage{xparse,xcolor}

\ExplSyntaxOn

\NewDocumentCommand{\setupbibcolors}{m}
 {
  \cs_set_protected:Npn \bibitem ##1
   {
    \color{ \str_case:nnF { ##1 } { #1 } { black } }
    \heba_bibitem:n { ##1 }
   }
 }

\cs_set_eq:NN \heba_bibitem:n \bibitem

\ExplSyntaxOff

\setupbibcolors{
  {baza1}{white}
  {baza2}{white}
  {baza3}{white}
  {baza4}{white}
   {baza5}{white}
  {baza6}{white}
  {baza7}{white}
  {baza8}{white}
   {baza9}{white}
    {baza10}{white}
    {baza11}{white}
    {baza12}{white}
    {baza13}{white}
     {baza14}{white}
}

\usepackage[ruled,linesnumbered]{algorithm2e}
\definecolor{seagreen}{rgb}{0.18, 0.55, 0.24}
\SetAlFnt{\small\color{black}\tt}
\LinesNumbered

\begin{document}
	\title{Privacy-preserving and Collusion-Resistant Charging Coordination Schemes for Smart Grids}

	\author{Mohamed Baza,Marbin Pazos-Revilla, Mahmoud~Nabil, Ahmed Sherif~\IEEEmembership{Member~IEEE},\\~Mohamed~Mahmoud~\IEEEmembership{Member~IEEE},~and~Waleed Alasmary~\IEEEmembership{Member,~IEEE,}
	\IEEEcompsocitemizethanks{\IEEEcompsocthanksitem Mohamed Baza and Mohamed Mahmud are with the Department of Electrical \& Computer
	Engineering, Tennessee Tech University, Cookeville, TN 38505 USA.\protect\\
		E-mails:\enskip{}\href{mailto:mibaza42@students.tntech.edu}{mibaza42@students.tntech.edu},
 \href{mmahmoud@tntech.edu}{mmahmoud@tntech.edu}.

	\IEEEcompsocthanksitem Marbin Pazos-Revilla is with Department of Computer Science and Engineering, University of South Florida, Tampa, FL, USA
	E-mail:\enskip{}\href{mailto:marbin@usf.edu}
	{marbin@usf.edu}

	\IEEEcompsocthanksitem Mahmoud Nabil is is with the Department of Electrical \& Computer Engineering, North Carolina A\&T University, Greensboro, NC, 27401 USA.
	E-mail:\enskip{}\href{mailto:mnmahmoud@ncat.edu}
	{mnmahmoud@ncat.edu}

	\IEEEcompsocthanksitem Ahmed Sherif is with School of Computing Sciences and Computer
Engineering, University of Southern Mississippi, USA.
	E-mail:\enskip{}\href{mailto:ahmed.sherif@usm.edu}
	{ahmed.sherif@usm.edu}
	
	\IEEEcompsocthanksitem Waleed Alasmary is with Department of Computer Engineering, Umm Al-Qura University, Makkah, Saudi Arabia.
	E-mail:\enskip{}\href{mailto:wsasmary@uqu.edu.sa}
	{wsasmary@uqu.edu.sa}
	
	}% <-this % stops an unwanted space

	\thanks{Manuscript received November xx, 2018; revised November xx, 2018.}

	}
	
	\IEEEtitleabstractindextext{%
	\begin{abstract}

Energy storage units (ESUs), including electric vehicles and home batteries, enable attractive features in the future smart grid. In this paper, we propose centralized and decentralized privacy-preserving and collusion-resistant charging coordination schemes.  The centralized charging coordination (CCC) scheme is used in case there is a robust communication infrastructure that connects the ESUs to a charging coordinator (CC) run by the utility, whereas the decentralized charging coordination (DCC) scheme is useful in case of remote areas or isolated microgrids in which a robust communication to the utility is not available or considered costly. 
In the CCC scheme, each ESU should acquire anonymous and unlinkable tokens from the CC to authenticate their charging requests and send them to the CC via a local aggregator. 
Having done that, if the CC and the aggregator collude, they cannot identify senders of the charging requests. Moreover, by sending multiple charging requests with random time-to-complete-charging (TCC) and the battery state-of-charge (SoC) by each ESU that follow a truncated normal distribution (instead of only one request), the CC cannot link the charging requests data sent from the same ESU at different time slots to preserve privacy. After receiving the charging requests, the CC runs an optimization technique to compute charging schedules to maximize the amount of power delivered to the ESUs before the charging requests expire without exceeding the maximum charging capacity. 
In the DCC scheme, charging is coordinated in a distributed way using data aggaregation technique. The idea is that each ESU selects some ESUs as proxies, and share a secret mask with each proxy. Then, each ESU adds a mask to its charging request and encrypt it using homomorphic encryption, so that by aggregating all requests all masks are nullified and the total charging demand for each priority level is known so that each ESU can compute its charging schedule. Due to using the masking technique, DCC is secure against collusion. The results of extensive experiments and simulations confirm that our schemes are efficient, secure, and can preserve ESU owners' privacy.

	\end{abstract}
	
	\begin{IEEEkeywords}
Privacy-preservation, Energy storage units, Charging coordination, Collusion attacks, Smart grid.
	\end{IEEEkeywords}
	
	}

	\maketitle

	\IEEEdisplaynontitleabstractindextext

	% \fi
	\IEEEpeerreviewmaketitle

	\IEEEraisesectionheading{\section{Introduction}\label{sec:introduction}}


	\IEEEPARstart{E}{nergy} storage units (ESUs), including home batteries and electric vehicles (EVs), will play a major role in the future smart grid~\cite{wang2018coordinated}. They can store energy when there is a surplus in energy generation and inject energy to the grid when the demand is high to balance the energy demand and supply, which in turn enhances the power grid resilience~\cite{7564967}. Moreover, ESUs can also facilitate the use of renewable energy generators by storing the excess energy generated~\cite{yang2018distributed}. Moreover, ESUs can also help electricity consumers to reduce their electricity bills by charging from the grid during low-tariff periods and power the houses during high-tariff periods. However, despite these benefits, ESUs pose several challenges that should be addressed for smoothing their integration with the power grid~\cite{tang2016online}. 

The simultaneous uncoordinated charging of ESUs may result in lack of balance between the charging demand and the energy supply.
For example, after work hours, most of the EVs' owners usually return home and plug in their EVs to charge.
The uncoordinated charging may result in stressing the distribution system, causing instability to the grid~\cite{del2016smart}, and could lead to a power outage in severe cases. 
To avoid such consequences, there is a substantial need for a \textit{charging coordination mechanism}~\cite{yang2019novel}. Typically, in a charging coordination mechanism, ESUs need to send charging requests that have data such as the time-to-complete-charging (TCC), the battery state-of-charge (SoC), and the amount of required charging to a charging controller (CC). 
Then, these data can be used to compute priority indices (i.e., charging priority) so that ESUs with the highest priorities should charge first without exceeding the maximum charging capacity, while 
other ESUs charging is deferred to future time slot~\cite{7564967}. 
Unfortunately, the data that should be reported to the CC can reveal sensitive information about the ESUs owners such as the location of an EV's owner, when he/she returns home, whether he/she is on travel, etc. To the best of our knowledge, many schemes were presented in the literature to deal with coordinated ESUs charging issues~\cite{yang2019novel,zou2016efficient,del2016smart,6204213,ma2011decentralized,6839071,7054726}, \textit{but they do not take the privacy issue into consideration}.

In this paper, we propose two privacy-preserving and collusion-resistant charging coordination schemes:  
\textbf{C}entralized \textbf{C}harging \textbf{C}oordination (CCC) scheme, and \textbf{D}ecentralized \textbf{C}harging \textbf{C}oordination (DCC). The CCC scheme is designed to work in smart grids in which there is a robust communication infrastructure that connects the ESUs to the utility. In this case, ESUs are connected to the main grid and charging coordination should be performed at the grid operators' level. On the other hand, the DCC is designed to work on remote areas such as isolated microgrids (island mode). The isolated microgrids are not connected to the main grid due to their remoteness or failure to connect to the main grid. In the DCC, various sources of distributed energy generators including renewable energy sources are the only solution to meet the energy needs of the isolated microgrids consumers, and the microgrid should function autonomously~\cite{yang2018distributed}.

\begin{comment}
The CCC scheme is used if there is a robust communication infrastructure that connects the ESUs to the utility and in case of connected microgrids while the DCC scheme is used in cases of remote areas such as isolated microgrids (island mode).
Connected microgids are connected to the main grid and charging coordination should be done at the grid operators' level, 
but the isolated microgrids are not connected to the main grid due to their remoteness or failure to connect to the grid.
In the latter case, various sources of distributed energy generators including renewable energy sources are the only solution to meet the energy needs of the isolated microgrids' consumers, and the microgrid should function autonomously.
\end{comment}

In the CCC, each ESU should use anonymous and unlinkable tokens from the CC to authenticate its charging requests and send them to a local aggregator. The tokens are generated using partial blind signature so that even if the CC and the aggregator collude, they cannot expose sender's true identity. However, the CC can link charging requests in different time slots to the same sender ESU because ESUs send requests with linkable (i.e., close) TCC and SoC values in consecutive time slots. By linking charging requests of an ESU, the CC can collect information that can be used to identify the ESU owner and as a result the ESU's owner privacy can be violated. To prevent this linkability, each ESU sends multiple charging requests with random TCC and SoC (instead of one request) that follow a truncated normal distribution while maintaining the charging priority of the ESU. By doing so, the CC cannot link the charging requests sent from the same ESU in different time slots to preserve privacy and the charging priority of the ESU is maintained. Then, upon receiving charging requests, the CC prioritizes the requests by running an optimization technique to maximize the power delivered to the ESUs before the charging requests expire while not exceeding the maximum charging capacity. Note that, this paper focuses on the privacy and security issues, and other research works have already studied the optimization techniques for charging coordination~\cite{zou2016efficient,wang2018coordinated,yang2019novel,yang2018distributed,del2016smart,6204213,ma2011decentralized,6839071,7054726}. While these schemes can be used with our scheme, we used a modified version of optimization technique for knapsack problem for the purpose of evaluations~\cite{R}.  

In DCC scheme, one node (or more) is selected as an \textit{aggregating node}. Each ESU selects a set of ESUs called \textit{proxies} and share a mask with each one. Each ESU then adds the mask to their charging requests, encrypts them using homomorphic encryption, and sends them to the aggregating node. By aggregating all requests, masks are nullified and only the total charging demand for each priority level is known. By this way, the aggregating node cannot know the charging requests of the ESUs to preserve privacy. Then, the aggregating node broadcasts the aggregated charging demand. If the demand is higher than the total charging capacity, the ESUs that have higher priority should charge without exceeding the charging capacity.

Our main contributions and the challenges the paper aims to address can be summarized as follows.
    \begin{itemize}

\item We studied the probability of linkability attacks by the CC in the centralized scheme using SoC and/or TCC. The results indicate that the CC can use SoC/TCC to link charging requests to a specific ESU successfully which violates the ESU owner's privacy.

\item A privacy-preserving CCC scheme is proposed. Collusion attacks between the CC and the local aggregator/ESUs is mitigated by using anonymous and unlinkable tokens. Also, linkability attacks are mitigated, and thus the CC cannot learn whether two charging requests of different time-slots are from the same ESU or not.
\item A privacy-preserving DCC scheme is proposed. The charging coordination can be performed without the need for a central party that can access the individual ESUs data (TCC and SoC). The scheme is also secure against collusion attacks that aim to reveal an ESU data.

\item Extensive simulations and analysis are conducted to evaluate the proposed schemes. The results indicate that our schemes can coordinate charging activities while preserving ESUs owners' privacy and securing them against collusion attacks.

\end{itemize}

The rest of the paper is organized as follows. We describe the network and threat models, followed by the design goals of our schemes in Section~\ref{ref: Net and threat model}. In Section~\ref{Preliminaries}, we discuss preliminaries used by our schemes. Then, the centralized and decentralized schemes are presented in Sections~\ref{Sec centrallized} and~\ref{sec: decentralized} respectively. Detailed security and privacy analysis are provided in Section~\ref{Sec: security and privacy}. In Section~\ref{Sec:  charging evaulations}, we discuss performance evaluations for our schemes. Section~\ref{related} presents the related work. Finally, we give concluding remarks in Section~\ref{conclusion}.

\section{Network/Threat Models And Design Goals}
\label{ref: Net and threat model}
In this section, we present the considered network models
followed by the adversary and threat models, and then, we
introduce the design goals of our schemes.
\label{sec:model}

\subsection{Network Models}

\begin{figure}[!t]
\centering
\includegraphics[width=.4\textwidth]{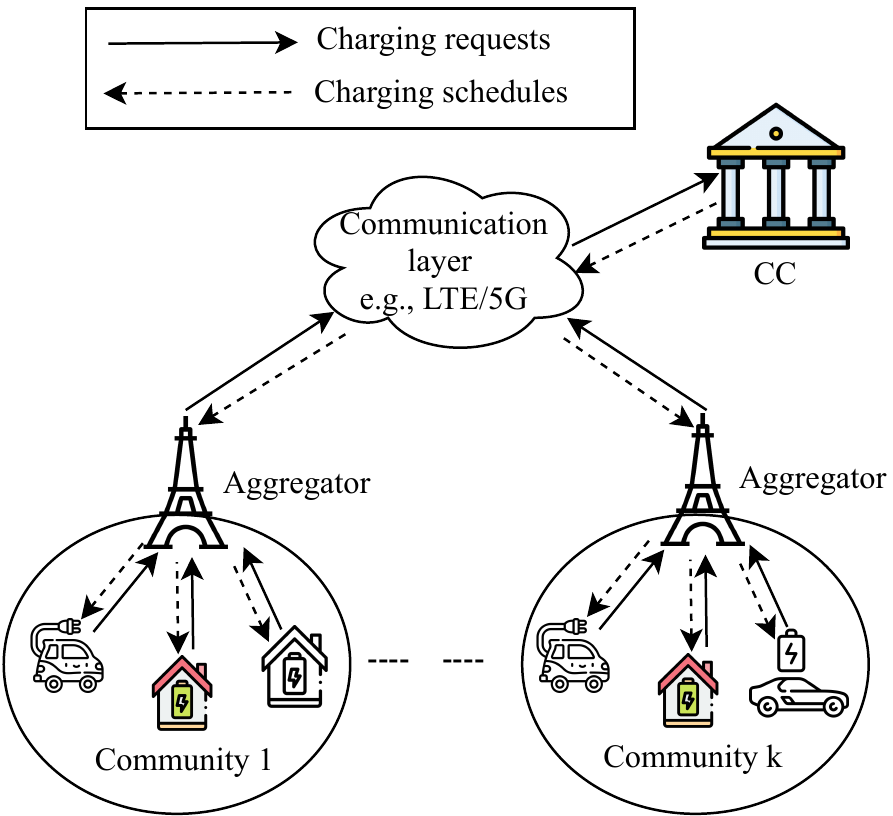}
\caption{Network model for the CCC.}  \label{fig centralized}
\bigbreak
\includegraphics[width=.43\textwidth]{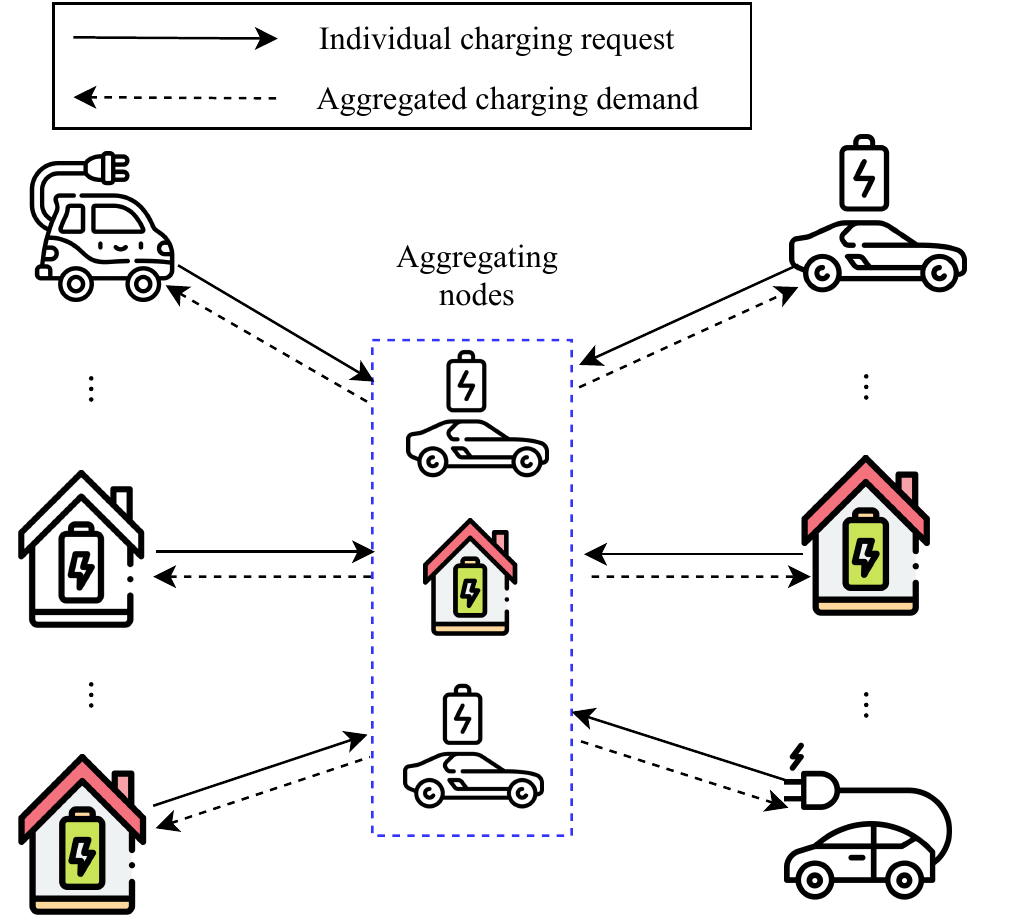}
\caption{Network model for the DCC.}\label{fig decentralized}
\end{figure}

As illustrated in Fig.~\ref{fig centralized}, the network model of the CCC scheme has a number of communities and CC. Each community has a group of ESUs and one aggregator. The storage units can be EVs or batteries installed in homes. The CC cannot communicate with
the storage units directly, but this has to be done via the aggregators.
The ESUs send charging requests to the aggregator to forward them to the CC. The CC prepares charging schedules and send them back to the ESUs via local aggregators.

As shown in Fig.~\ref{fig decentralized}, the network model of the DCC scheme has only ESUs that can communicate with other ESUs using one-hop or multihop communication protocols~\cite{mahmoud2013secure}. In multihop communication, some ESUs can act as routers to relay other ESUs messages. As shown in the figure, each ESU should send its individual charging demand to aggregating node (s) to aggreagte them and return aggregated charging demand. Then, each ESU can compute its priority index and use it to compute its charging schedules. An aggregating node ESU is responsible for receiving charging requests from  other ESUs, aggregate and decrypt them, and finally broadcast the aggregated message. The scheme can be run by one aggregating ESU that changes each time the scheme is run to distribute the computation overhead on the ESUs. Multiple (or all) ESUs can also act as aggregating nodes to ensure that the aggregation of the charging demands are done properly.

\subsection{Adversary and Threat Models}

In both CCC and DCC,  the attackers are interested in gathering private information about ESUs owners. These information include whether the owner is on travel, when he/she returns home, and other daily activities.

The attackers in the CCC scheme can be the aggregator, the CC, ESUs, or external eavesdroppers who may passively snoop on the communications to learn sensitive information. Attackers can work individually or collude to launch stronger attacks to reveal SoC/TCC of specific ESUs. Also, the CC can launch a linkability attack using the SoC and/or TCC of charging requests to infer an ESU sensitive information.

In the DCC scheme, the attackers can be ESUs including aggregating nodes, or eavesdroppers that eavesdrop on the communications of the ESUs and try to figure out some sensitive information on ESUs' owners. The attackers can work individually or collude to launch stronger attacks.

\subsection{Design Goals}

Our schemes should achieve the following important features.

\begin{itemize}

\item \textit{Privacy-preserving charging coordination}. A charging coordination scheme should compute charging schedules that charge the highly prioritized requests and deter other requests to future time slots without exceeding the maximum charging capacity, and without revealing any sensitive information about ESUs owners.

\item \textit{Resistance to collusion attacks.} Our schemes should be secure against collusion attacks, i.e., if any collude, they should not obtain SoC/TCC of a victim ESU.

\item \textit{Resistance to linkability attacks}. No one including the CC and the aggregating nodes should be able to link charging requests and the corresponding SoC/TCC sent from an ESU at different time slots.

\item \textit{Data integrity and authenticity.} The integrity and authenticity of the charging requests should be verified.

\end{itemize}

\begin{table}[!t]
  		\renewcommand{\arraystretch}{.8}
  	\caption{Notations}
  	\label{not}
  	\scalebox{0.9}{
  	\begin{tabular}{m{2.7cm}||m{5.8cm}}
  	 \Xhline{3.4\arrayrulewidth}  \\[5pt]
  		Notation & Description \\[6pt]
  		\hline\hline\\
  
  	$\mathcal{B}(m)$ & Blined message $m$ \\[6pt]
  		$\sigma^{c}_{d}(\mathcal{B}(m))$ & Partially blinded signature on message $m$ where $c$ is the common appended message, and $d$ is the signer's private key. \\[8pt]
  		$K_{v\leftrightarrow cc}$& Shared symmetric key between an ESU $v$ and the CC   \\[8pt]
  		
  		$U_v$ & The charging priority of an ESU $v$\\[6pt]
  		$P_{CC}, S_{cc}$ &   The CC public/private key pair  \\[6pt]
  		$\tau^{(\ell)}_{v}$ & A charging token $\ell$ of an ESU $v$\\[6pt]
  		
  		$PK^{(\ell)}_{v}/SK^{(\ell)}_{v}$ &A public/private key pair for each token \\[6pt]
  		
                	$S_{v}$, 	$\mathcal{T}_{v}$& SoC/TCC of an ESU $v$  \\[6pt]

                $R^{(j)}_{v}$ & A charging request $j$ from an ESU $v$	\\[4pt]

			$R^{(j,k)}_{v}$ & Individual $n$ charging requests of the request $R^{(j)}_{v}$, $1\leq k\leq n$	\\[4pt]
			
			$S^{(j,k)}_{v}$, $\mathcal{T}^{(j,k)}_{v}$, $U^{(j,k)}_{v}$ & SoC, TCC, priority for	$R^{(j,k)}_{v}$ \\[4pt]

$Sc^{(j,k)}_{v}$ & Charging schedule of an ESU $v$ at a time slot $j$ \\[4pt]
  		$q, G_1, G_2, P, e$  & Public parameters of bilinear pairing\\[6pt]
  		
  		 $(N, g)$, $\delta$ & Public/private key of homomorphic encryption\\[6pt]

				$v_A$ & Aggregating node \\ \Xhline{3.4\arrayrulewidth}   
  		
  	\end{tabular}}
  \end{table}

\section{Preliminaries}

\label{Preliminaries}

In this section, we present the necessary background on bilinear pairing, partial blind signature, and paillier cryptosystem that will be used in this paper. Notations used in the paper are given in Table.~\ref{not}.

\subsection{Bilinear Pairing}

Let $G_1$ be a cyclic additive group with generator $P$ and order of prime $q$, and $G_2$ be a cyclic multiplicative group with the same order. Let $e$: $G_1 \times G_1$ $\rightarrow$ $G_2$ be a bilinear map with the following properties. 
\begin{itemize}
\item \textit{Bilinearity}: $e(aP, bQ$)  = $e(P,Q)^{ab}$, where $P, Q \in$ $G_1$, and $a$, $b$ $\in$ $Z^*_q$. 
\item \textit {Non-degeneracy}: There exists $P,Q$ $\in$ $G_1$ such that $e(P,Q$) $\neq$ 1. 
\item \textit{Computability}: There is an efficient algorithm to compute $e$($P,Q$) for all $P,Q$ $\in$ $G_1$.
\end{itemize}
\subsection{Partial Blind Signature}

\label{PBS}

Blind signature is a cryptosystem in which a sender of a message is able to get the signature on this message from the singing party while concealing the content of the message. Partial blind signature (PBS) is a special case of blind signature in which the signer can include information that is known to both singer and sender in the signed message, such as a time or date \cite{Okamoto2006}. A brief description of the PBS proposed in~\cite{zhang2003efficient} is given as follows. 

\begin{enumerate}
\item The signer picks an element $d$ $\in_R$ $Z^*_q$ as the private key and computes public key $P_{pub}$ = $dP$, where ${P}$ is the generator of a cyclic additive group ${G_1}$.

\item The requester randomly chooses a number $r \in_R Z^*_q$ and computes $\mathcal{B}(m) = \mathcal{H}_o(m\vert\vert c) + r(\mathcal{H}(c)P + P_{pub}$), where $\mathcal{B}(m)$ is the blinded message $m$, $\mathcal{H}$ is hash function such that $\mathcal{H}_0$: \{0,1\}$^*$ $\rightarrow$ $Z^*_q$, $\mathcal{H}_0$ is hash function such that $\mathcal{H}$: \{0,1\}$^*$ $\rightarrow$ $G_1$, and $c$ is the common information e.g., date. Then, the requester sends $\mathcal{B}(m)$ to the signer.

\item The signer sends back $\sigma^{c}_{d}(\mathcal{B}(m))= (\mathcal{H}(c) + d)^{-1}\mathcal{B}(m)$ to the requester.  

\item The requester applies unblinded operation $\mathcal{B}^{-1}$ using the secret key $r$ to $\sigma^{c}_{d}(\mathcal{B}(m))$ to obtain the signer's signature  $\sigma^{c}_{d}(m)$ as follows.

\begin{align}
\mathcal{B}^{-1}\Big(\sigma^{c}_{d}(\mathcal{B}(m))\Big) &= \sigma^{c}_{d}(\mathcal{B}(m)) - rP  \nonumber \\ &  = \frac{\mathcal{H}_o(m\|c)}{\mathcal{H}(c)+d} = \sigma^{c}_{d}(m)
\label{eq: PBS sig}
\end{align}

\end{enumerate}

Finally, the requester can use $m \| \sigma^{c}_{d}(m)$ to authenticate itself anonymously and the signer can accept the signature by checking:

$$e(\mathcal{H}(c)P+P_{pub}, \sigma^{c}_{d}(m)) \stackrel{?}{=}e(P, \mathcal{H}_o(m\vert\vert c))$$

\subsection{Paillier Cryptosystem}

Paillier cryptosystem~\cite{Pailler1999homomorphic} is one of the popular techniques to achieve homomorphic additive encryption. In Paillier cryptosystem, if two messages $m_1$ and $m_2$ are encrypted as $E_{k}(m_1)$ and $E_{k}(m_2)$ with the same key $k$, to obtain the ciphertext of the summation of $m_1$ and $m_2$, the two ciphertexts of $m_1$ and $m_2$ are multiplied, i.e.,
$$
E_{k}(m_1) \cdot E_{k}(m_2)=E_{k}(m_1+m_2)
$$
Typically, Paillier cryptosystem is composed of the following
phases: key generation, encryption, and decryption.

\begin{enumerate}
    \item \textit{Key Generation:}  two large and independent prime
numbers $p$ and $q$ are selected randomly, and $\delta=\operatorname{lcm}(p-1, q-1 )$ and $N=p \cdot q$ are computed where $\delta$ is the least common multiple of $p-1$ and $q-1$. Then, a function $L(x)=\frac{x-1}{N}$ is defined, a generator $g=(1+N)$ is chosen, and $\mu=\left(L\left(g^{\delta} \bmod N^{2}\right)\right)^{-1} \bmod N$ is computed. The public
key is $(N, g)$, and the private key is $(\delta, \mu)$.

\item \textit{Encryption:} Given message $m \in \mathbb{Z}_{N}^{*}$, and a random number $r \in Z_{N^{2}}^{*}$, the ciphertext can be computed as follows:

$$
\mathcal{C}=E(m)=g^{m} \cdot r^{N} \quad \bmod N^{2}
$$

\item \textit{Decryption:} 
Given a ciphertext $\mathcal{C}$, where
$\mathcal{C} \in \mathbb{Z}_{N^{2}}^{*}$, we can compute the plaintext message as
$$
m=L\left(\mathcal{C}^{\delta}  \bmod N^{2}\right) \cdot \mu \quad \bmod N
$$

\end{enumerate}

\section{The CCC Scheme}
\label{Sec centrallized}

In this section, we discuss and evaluate charging request linkability attacks, and then we discuss in details our CCC scheme.

\subsection{Linkability Attacks}
\label{sec motivation}

In the CCC scheme, the CC should access the SoC and TCC of ESUs' charging requests to compute the charging schedules. However, since the CC has these data over several time slots for all the ESUs, it could link charging requests sent from the same ESU, which would violate the privacy of the ESUs's owners. This attack is known as a \textit{linkability attack}. Typically, the CC can link an ESU charging requests because the SoC/TCC of the ESU requests are related. For example, if an ESU does not charge in a time slot, the SoC of the next time slot should at most equal by the energy consumption amount. Therefore, the two SoCs can be linked by approximating the consumption amount and finding the request that has the closest SoC value.

To show the effectiveness of the linkability attack based only on the SoC/TCC values, we use Matlab to simulate the attacks and run experiments. The simulation parameters are set as follows. The maximum charging capacity of a community is 1000 KW, the maximum charging capacity of an ESU was set to 100 KW, and the number of ESUs is 150. TCC and SoC are selected randomly based on a uniform distribution from $\{1,\cdots,48\}$ in time slots and $[1, 50]$ KW respectively. Three attacks are simulated to link charging requests to the same ESU. \textit{Attack 1} uses only SoC to link requests, \textit{Attack 2} uses only TCC, and \textit{Attack 3} uses both SoC and TCC. Two cases are simulated in all attacks including low-resolution and high-resolution values for SoC. For the low-resolution case, the change in reported SoC is relatively small (i.e., takes integer values) in consecutive time slots. For the high-resolution case, the change in SoC takes the range of decimal values. Also, we define the probability of a successful linkabilty attack as the ratio between the successfully linked ESUs between time slot $j$ and time slot $j+1$ to the total number of ESUs at time slot $j$. For example in case of \textit{Attack 1}, two ESUs at time slot $j$ and $j+1$ are said to be linked if the Euclidean distance between their respective SoC in the two consecutive time slots is less than certain threshold (i.e., the average energy consumption of an ESU in a time slot). Similarly, TTC or a combination between SoC and TCC can be used to determine the probability of a successful linkabilty attack. In all experiments, 100 runs were performed, and the average values were taken.

\begin{figure}[!t]
\centering
\includegraphics[width=.47\textwidth]{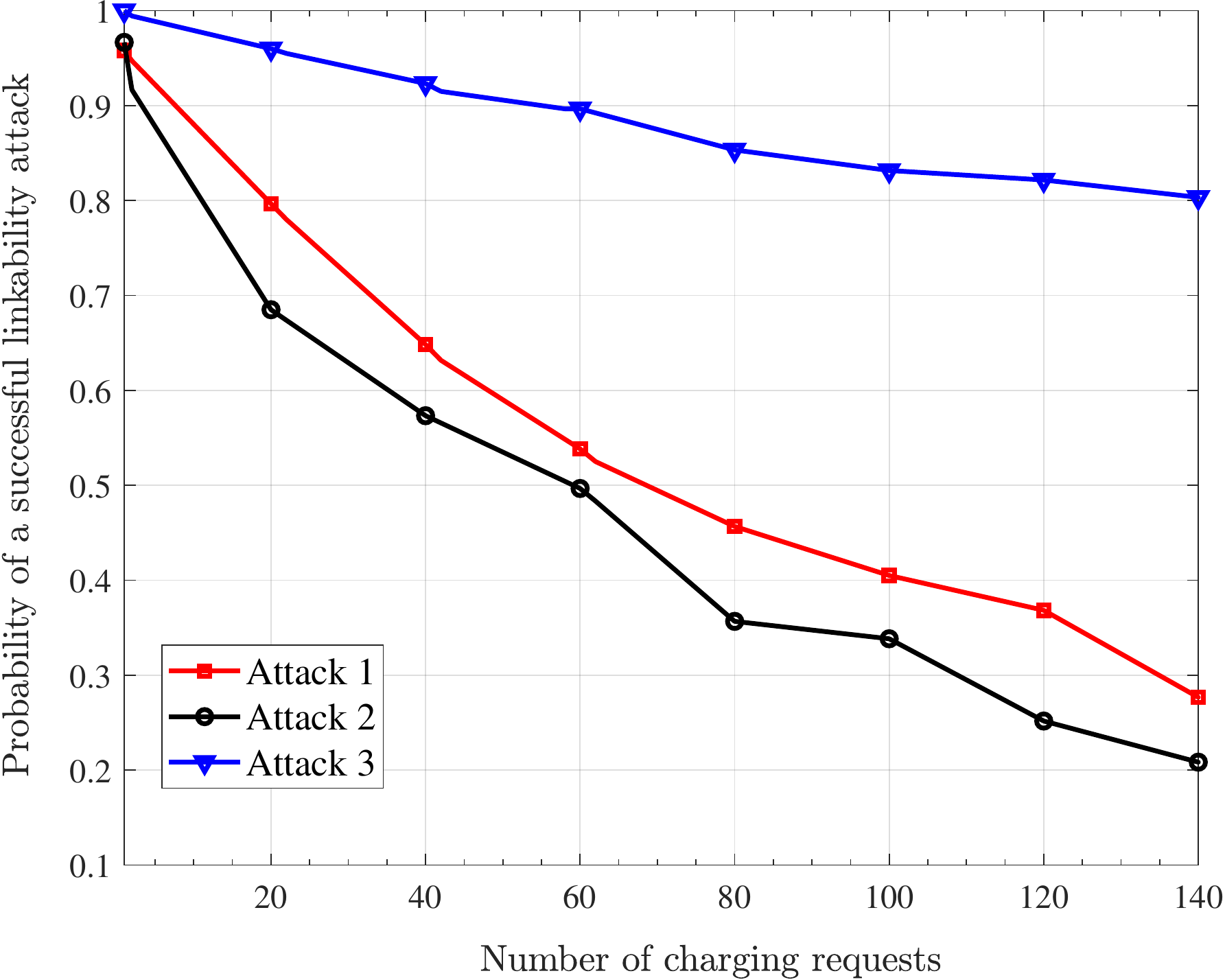}
\caption{The probability of successful linkability attacks with low resolution of SoC.}  \label{fig:link attacks a}
\bigbreak
\includegraphics[width=.47\textwidth]{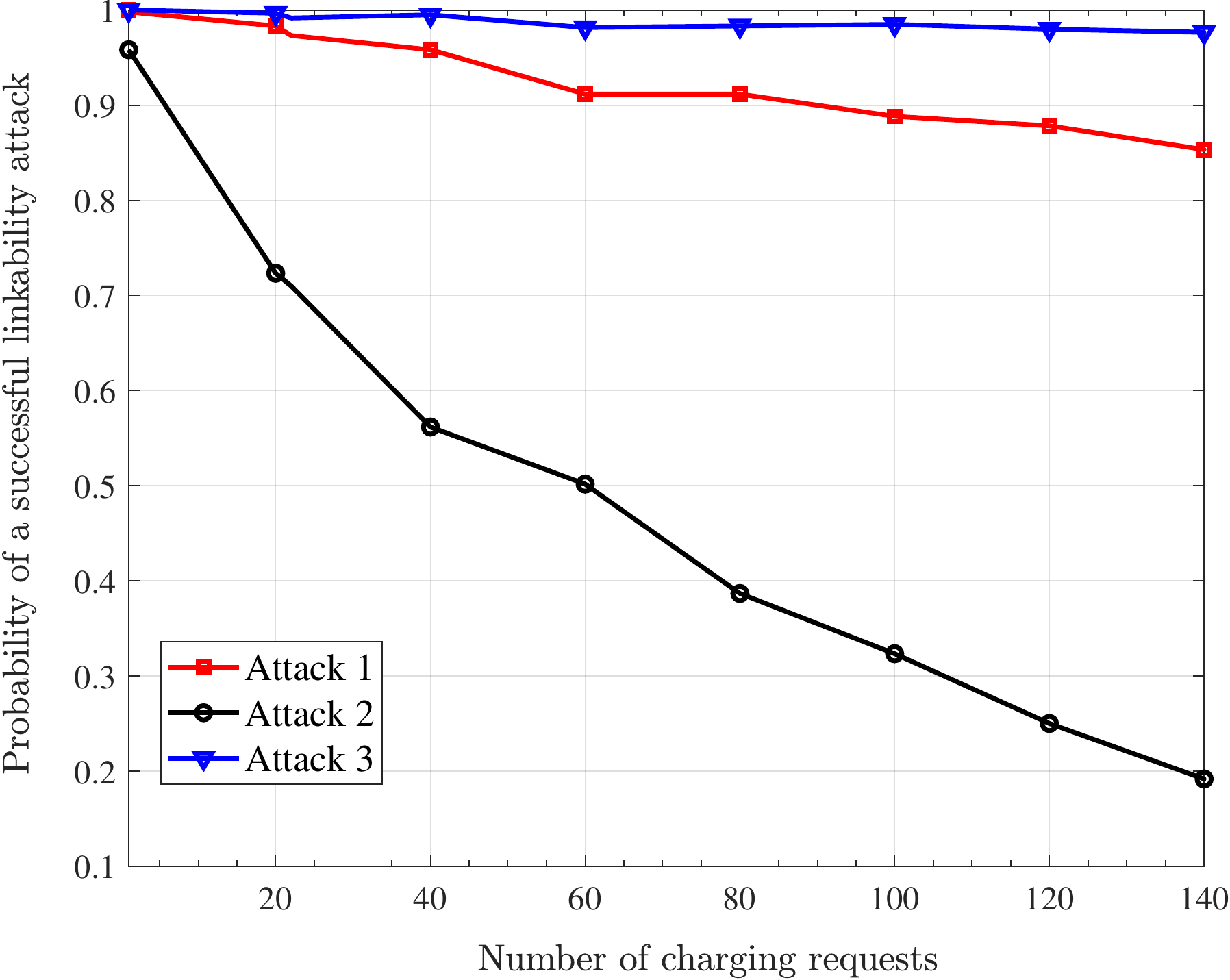}
\caption{The probability of successful linkability attacks with high resolution of SoC.}\label{fig:link attacks b}
\end{figure}

As shown in Fig.~\ref{fig:link attacks a}, for a low resolution of SoC, charging request linkability attack can experience considerable levels of success. For instance, the average probability of success is above 0.75, 0.65, and 0.95 for \textit{Attack 1}, \textit{Attack 2} and \textit{Attack 3}, respectively in case of 20 ESUs. While it reaches 0.2 in the case of \textit{Attack 1} when the number of ESUs increases to 140, or below 0.2 in the case of \textit{Attack 2}, however, when the combination of SoC and TCC is used in \textit{Attack 3}, the likelihood of success is considerably higher, reaching values of over 0.8. This is because as more ESUs submit requests, the likelihood of similarity among requests increases, which explains the drop in the success rate of linkability attacks as the number of ESUs increases.

For high-resolution values of SoC, it can be clearly seen from Fig.~\ref{fig:link attacks b} that the success rate of the attacks increases considerably compared to the low-resolution SoC cases given in Fig. \ref{fig:link attacks a}. This is noticeable especially for \textit{Attack 3}, where the success probability reaches 0.97 even with a high number of ESUs as 140. In the case of \textit{Attack 2}, the success probability is not significantly affected compared to Fig.~\ref{fig:link attacks a}. This behaviour is attributed to smaller values of TCC comparing to those of SoC. With a larger set of possible values for SoC, it becomes easier to single out distinct pairs of requests. This contributed to the increased level of success of the likability attack, either when SoC is used by itself, or in combination with TCC.

It can be concluded that \textit{Attack 1} is always more successful than \textit{Attack 2} because SoC has a larger range of values than TCC. This makes it easier for the CC to link requests sent from an ESU as there is more range of values each request can take. Also, \textit{Attack 3} is always more successful than \textit{Attacks 1} and \textit{Attack 2} because \textit{Attack 3} can benefit from the range of values of SoC and the additional information of TCC that can contribute to the success of the attack. Moreover, the results of the above experiments demonstrate that the number of requests submitted from a community would need to be sufficiently large in order to make data linkability attack unsuccessful. Motivated by these results, in our scheme, each ESU sends multiple charging requests (instead of only one) and by properly selecting the SoC/TCC of the ESUs' requests, these requests should have close values of SoC/TCC to make linkability unsuccessful but SoC/TCC should also maintain the priority level of the ESU.

\begin{figure} [!t]

 \begin{center}

\includegraphics[width=0.38\textwidth]{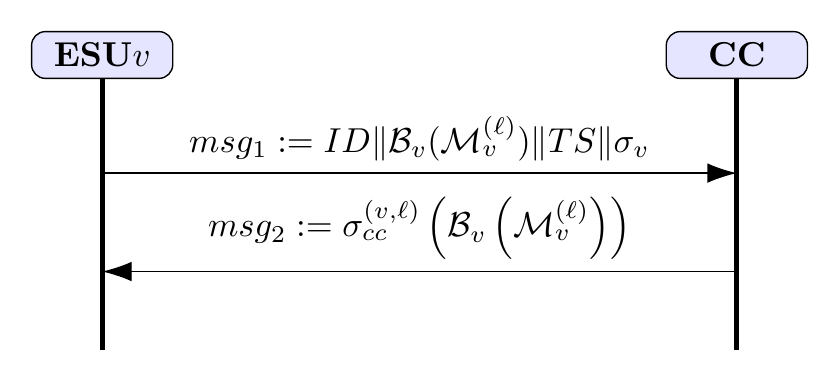}

        \caption{Token acquisition.}
\label{fig:proposed}
\end{center}
\end{figure}

\begin{comment}
\begin{figure} [!t]

 \begin{center}

\includegraphics[width=0.499\textwidth]{Figs/97.pdf}
\hspace{0.4mm}
        \caption{Exchanged messages in the CCC scheme.}
\label{fig:proposed}
\end{center}
\end{figure}
\end{comment}

\subsection{Overview of the CCC scheme}
The CCC scheme consists of four phases. First, in the \textit{acquisition of tokens phase}, an ESU acquires a number of cryptographic tokens from the CC to anonymously authenticate itself to the CC. Then, in the \textit{charging requests submission phase}, ESUs use the tokens to send their charging requests to the local aggregator. Then, for efficiency, in the \textit{verification of charging requests and signature aggregation phase}, the aggregator should verify all the received signatures, aggregate them, and send one signature to the CC. Finally, in the \textit{computing charging schedules phase}, the CC compute the charging schedules and return them back to the aggregators.

\subsection{Acquisition of Tokens}
\label{sec tokebns}

In this phase, each ESU acquires a number of cryptographic tokens from the CC. These tokens are used to anonymously authenticate the ESU and also help it to share a key with the CC to encrypt the charging schedules. Acquisition of tokens is illustrated in Fig.~\ref{fig:proposed}.

Assume that the CC has a public/private key pair $\{P_{CC}, S_{cc}\}$. Each ESU $v$ acquires $m$ tokens. For each token, $v$ generates a public/private key pair $PK^{(\ell)}_{v}/SK^{(\ell)}_{v}$, where $1 \leq \ell \leq  m$, and computes $[K^{(\ell)}_{v\leftrightarrow cc}]_{P_{cc}}$ which denotes a symmetric key $K^{(\ell)}_{v\leftrightarrow cc}$ encrypted by the CC public key (${P_{cc}}$) 

Then, $v$ should send $msg_1$ to the CC that is: 
$$msg_1:=ID\|\mathcal{B}_{v}({\mathcal{M}^{(\ell)}_{v}}) \| TS \| \sigma_{v}, $$
 
where $ID$ is the ESU real identity, $\mathcal{M}^{(\ell)}_{v}=PK^{(\ell)}_{v}\| [K^{(\ell)}_{v\leftrightarrow cc}]_{PK_{cc}}$, $\mathcal{B}_{v}({\mathcal{M}^{(\ell)}_{v}})$ is the blinded message of $\mathcal{M}^{(\ell)}_{v}$, ${TS}$ is the time-stamp, and $\sigma_{v}$ is the signature on the whole message. Note that $\sigma_{v}$ reveals the associated $v$ real identity and appending the time-stamp as part of the request to protect against packet \textit{replay} attack.

Then, once the CC receives $msg_1$, it verifies the authenticity of the request by verifying the signature $\sigma_{v}$. In addition, it checks that the request time-stamp matches the current time. If all the verifications succeed, the CC signs the request and sends a partially-blind-signature $msg_2$ back to the ESU as follows

$$msg_2:=\sigma^{(v,\ell)}_{cc}\left(\mathcal{B}_{v}\left({\mathcal{M}^{(\ell)}_{v}}\right)\right).$$

Note that appended common message of the PBS $c=TE\|ID_g$ where $TE$ is the token's expiry date and $(ID_g)$ is the identifier of the community. Then, $v$ applies unblinded operation $\mathcal{B}^{-1}_{v}$ to obtain the signature on the token $\mathcal{M}^{(\ell)}_{v}$ as follows

%$$\mathcal{B}^{-1}_v\left(\mathcal{PBS}^{m_0}_{S_{cc}}\left(\mathcal{B}_{v}\left({\tau^{(\ell)}_{v}}\right)\right)\right)= \mathcal{PBS}^{m_0}_{S_\tau}\left({\tau^{(\ell)}_{v}}\right)$$

$$\mathcal{B}^{-1}_v\left(\sigma^{(v,\ell)}_{cc}\left(\mathcal{B}_{v}\left({\mathcal{M}^{(\ell)}_{v}}\right)\right)\right)= \sigma^{(v,\ell)}_{cc}\left({\mathcal{M}^{(\ell)}_{v}}\right)$$

Finally, $v$ uses the following token denoted as $\tau^{(\ell)}_{v}$  

$$\tau^{(\ell)}_{v}=\mathcal{M}^{(\ell)}_{v}\|\sigma^{(v,\ell)}_{cc}\left({\mathcal{M}^{(\ell)}_{v}}\right),$$

to authenticate itself anonymously to its local aggregator and the CC.

\begin{figure} [!t]
 \begin{center}
\includegraphics[width=0.5\textwidth]{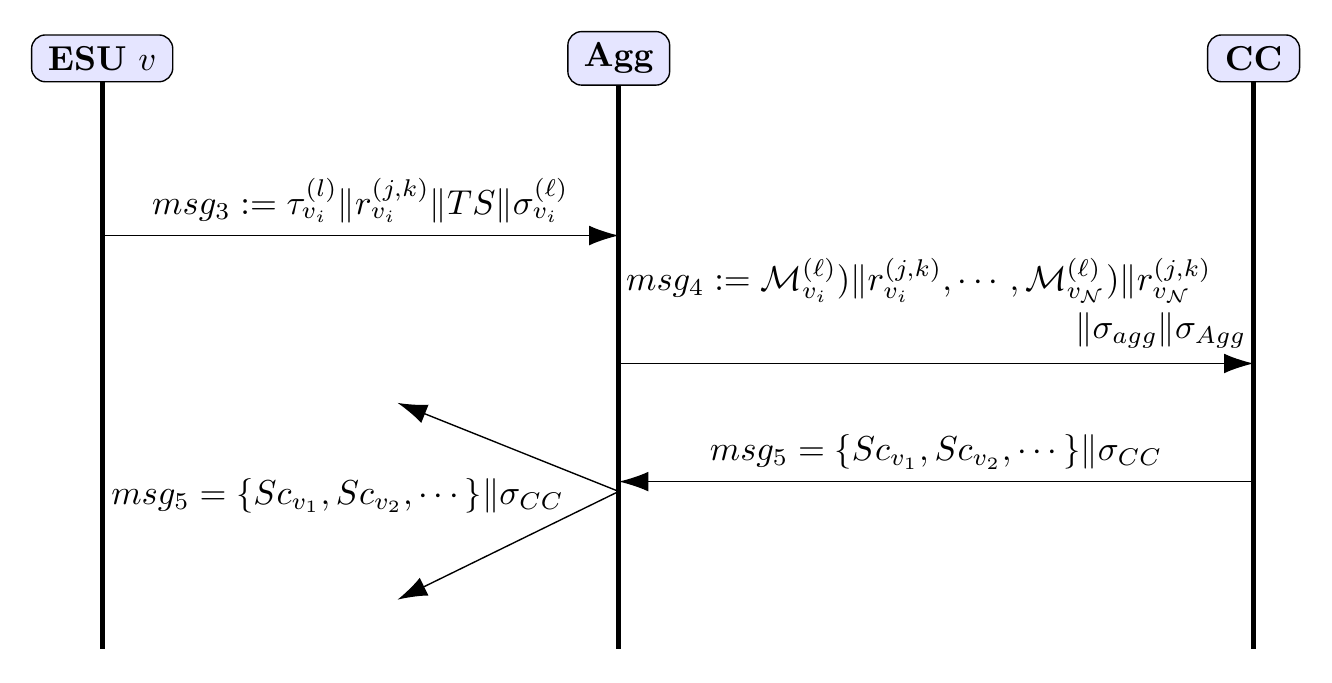}
        \caption{Charging request submission and charging schedules.}
\label{fig:proposed1}
\end{center}
\end{figure}

\begin{figure*}[!ht]
	\setlength{\abovecaptionskip}{0.1cm}
	\setlength{\belowcaptionskip}{-0.7cm}
	\centering
	\subfloat[at $s$ = 0.4.\label{fig:priorty a}]
    {\includegraphics[width=0.3333333333\linewidth]{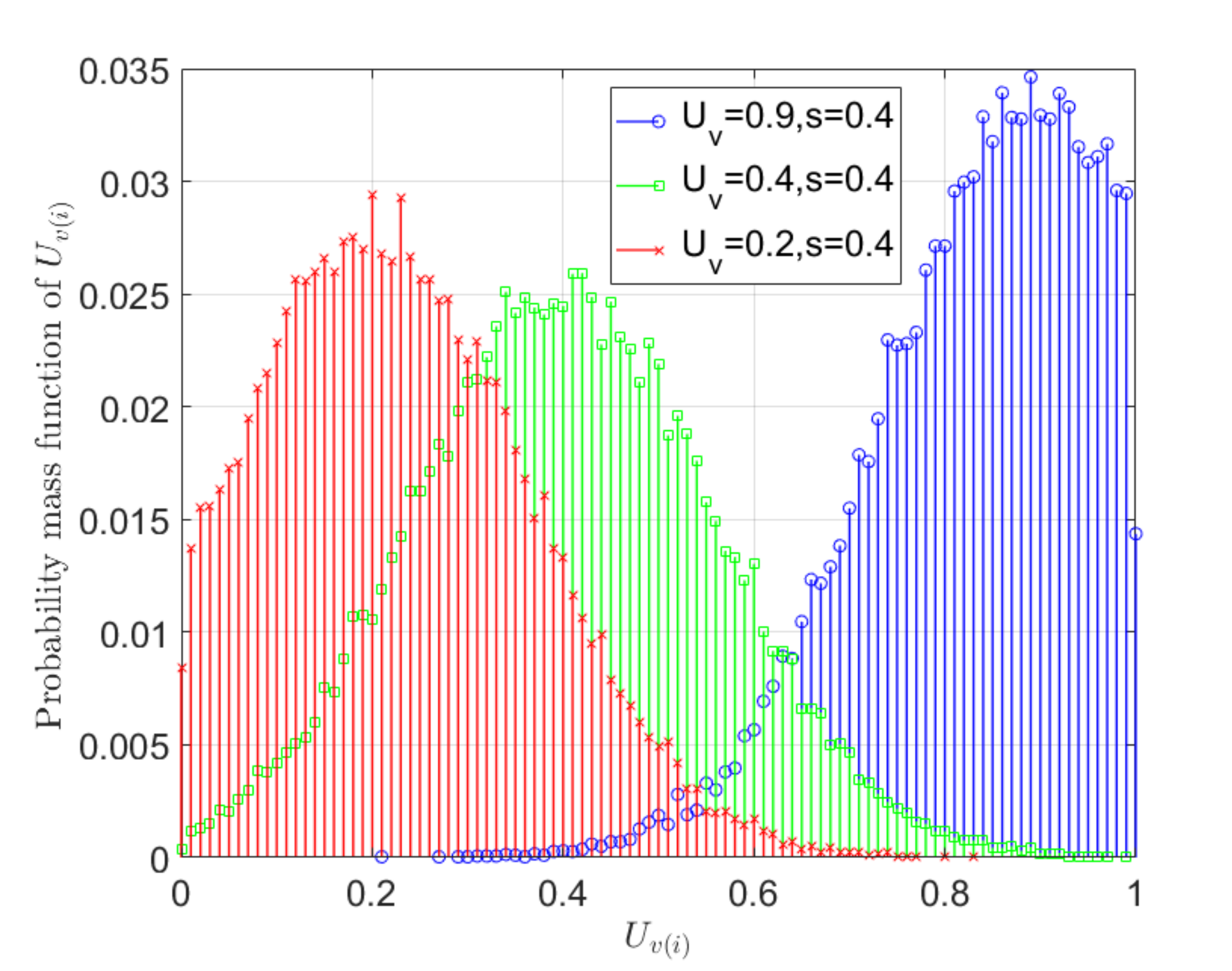}}
    \subfloat[at $s$ = 0.05.\label{fig:priorty b}]
	{\includegraphics[width=0.3333333333\linewidth]{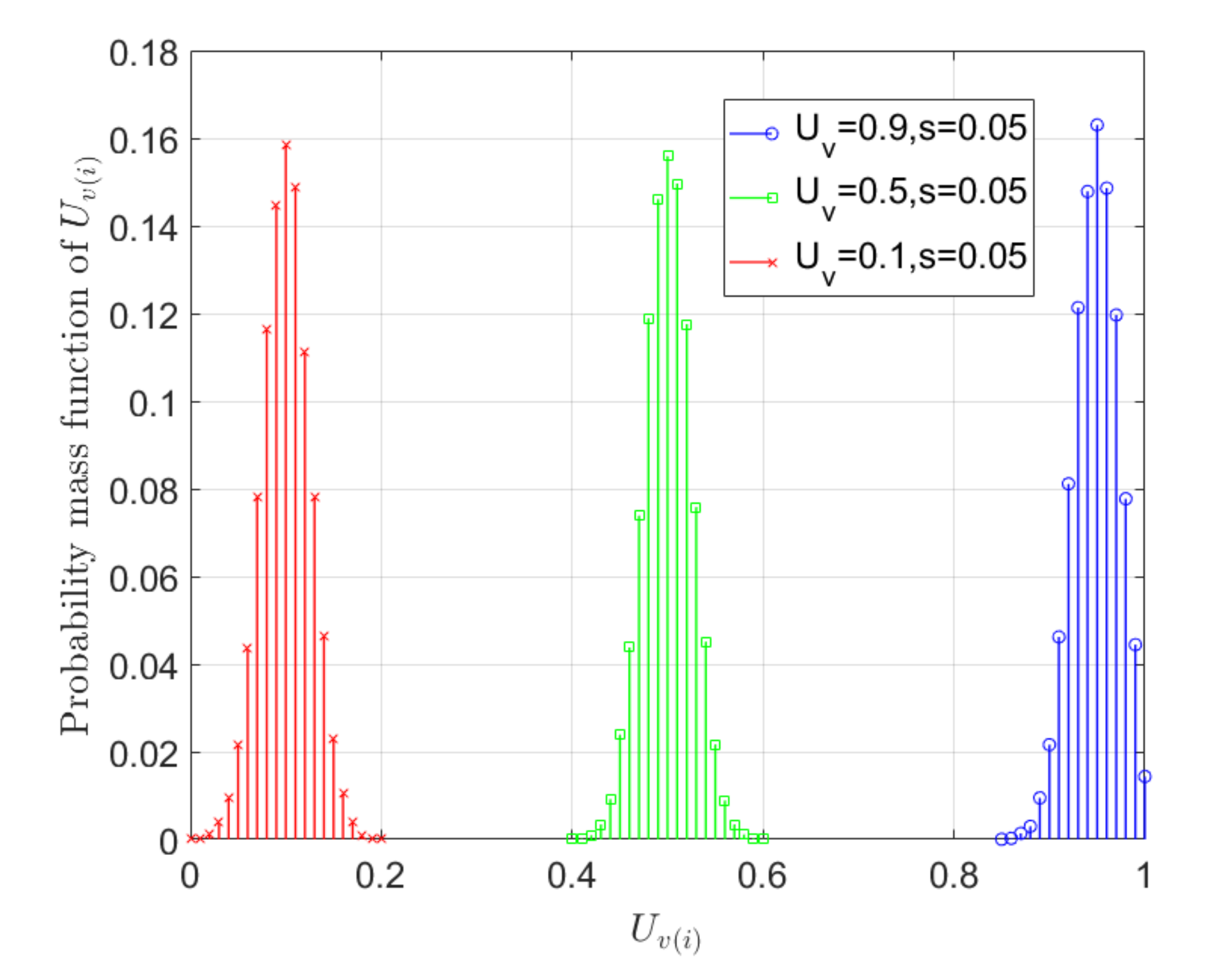}}
    \subfloat[at $s$ = 0.85. \label{fig:priorty c}]
	{\includegraphics[width=0.3333333333\linewidth]{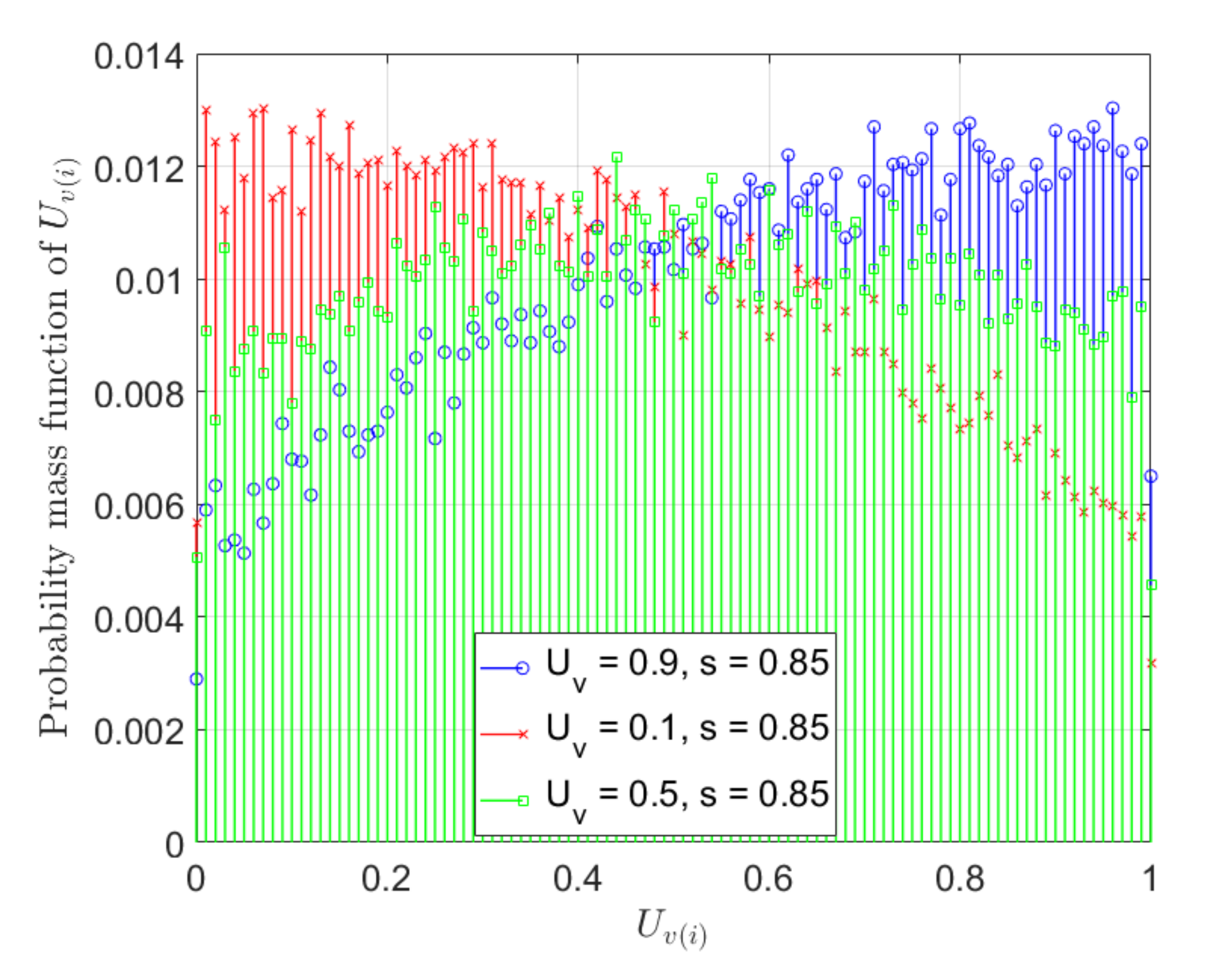}}
	\vspace{10pt}
	\caption{Probability mass function of ESU priority at different values of the variance ($s$).}
	\label{fig:priorty}
	\vspace{10pt}
\end{figure*}

\subsection{Charging Requests Submission}
\label{charging_request}

In this subsection, we discuss how an ESU sends their charging requests to the CC anonymously while mitigating linkability attacks. Exchanged messages of charging request submission and scheduling phases are illustrated in Fig.~\ref{fig:proposed1}.

\vspace{7pt}
\subsubsection{Computing SoC/TCC of Charging Requests} 
The priority of an ESU $v$ in a given time-slot can be mathematically expressed using its SoC and TCC as follows.

\begin{equation}
    U_v = \alpha_1 (1 - S_v) + \alpha_2 F(\mathcal{T}_v),
     \label{1}
\end{equation}

Where $F(\mathcal{T}_v)$ is a decreasing function of TCC $(\mathcal{T}_v)$ with a range of $[0,1]$ such that $F(\mathcal{T}_v) = 0$ for long TCC and equals $1$ for short TCC, and SoC value ($S_v) \in [0,1]$ with $S_v = 1$ for a completely charged ESU. Both $\alpha_1$ and $\alpha_2$ are weights that are appropriately chosen to give relative importance for $S_v$ and $F(\mathcal{T}_v)$, with $\alpha_1$ + $\alpha_2$ = 1.

Our strategy to mitigate linkablity attacks is as follows. For a charging request denoted as $R^{(j)}_{v}$ where $j$ is the current time-slot, $v$ creates $n$ individual requests, and each request is denoted as $R^{(j,k)}_{v}$ where $(1\leq k \leq n)$. Each request $R^{(j,k)}_{v}$ should be sent independently during the same time slot. To calculate the SoC and TCC of each request, $v$ first uses its SoC  ${S^{(j)}_{v}}$ and TCC ${\mathcal{T}^{(j)}_{v}}$ at the time slot $j$ to compute a priority ${U^{(j)}_{v}}$ using Eq.~\ref{1}. Then, using ${U^{(j)}_{v}}$, it computes $n$ random priorities ${\{U^{(j,1)}_{v},\cdots, U^{(j,n)}_{v}\}}$ for each individual request $R^{(j,k)}_{v}$. Finally, for each individual priority $U^{(j,k)}_{v}$, it calculates random tuples of ${S^{(j,k)}_{v}}$ and ${\mathcal{T}^{(j,k)}_{v}}$ that can achieve the priority $U^{(j,k)}_{v}$ using Eq.~\ref{1}. 

\begin{comment}

To implement the idea, let the priority of an ESU $v$ is $U^{(j)}_{v}$ at a time slot $j$, and we want to get $n$ priorities $\{U^{(j,k)}_{v},\cdots,U^{(j,n)}_{v}\}$ such that they are; (1) close or equal to $U^{(j)}_{v}$; (2) random in each time; and (3) in range of [0,1] since $0\leq {{U_{v}}} \leq$ 1. Thus, to achieve previous requirements mathematically, we can sample from a PDF of a truncated normal distribution~\cite{burkardt2014truncated} for the following reasons. Firstly, most probably the values in truncated normal distribution are close to the distribution mean (requirement (1)). Secondly, sampling means randomness (requirement (2)). Thirdly, it is bounded from $a$ and $b$ which can be mapped to [0, 1] (requirement (3)). Accordingly, the PDF of the truncated normal distribution can given by~\cite{burkardt2014truncated}.
\end{comment}

To implement the idea, let the ESU priority $U^{(j)}_{v}$ at a time slot $j$. To generate $n$ individual priorities $\{U^{(j,1)}_{v},\cdots,U^{(j,n)}_{v}\}$, the following conditions need to be satisfied for each individual priority $U^{(j,k)}_{v}$:
\begin{itemize}[label={}]
    \item \textbf{C.1:} It needs to be random or have different values in order to mitigate the linkability attacks.
\item \textbf{C.2:} It needs to be close or equal to $U^{(j)}_{v}$. This to ensure that each individual request $R^{(j,k)}_{v}$ has the same importance to be served as the original request $R^{(j)}_{v}$ so that the charging performance is not affected.

\item \textbf{C.3:} It needs to be in the range [0,1] as we deal with it as a probabilistic value.
\end{itemize}

Thus, to achieve previous requirements mathematically, each individual priority $U^{(j)}_{v}$ can be sampled from the PDF of a truncated normal distribution~\cite{burkardt2014truncated} for the following reasons. Firstly, by sampling, a random value is selected from the PDF of the truncated normal distribution (\textbf{C.1}). Secondly, most probably the values in truncated normal distribution are close to the distribution mean (\textbf{C.2}). Thirdly, it is bounded from $a$ and $b$ which can be mapped to [0, 1] (\textbf{C.3}). Accordingly, the PDF of the truncated normal distribution with variance $s$ and mean $\mu$ can be used. Details on how an ESU can sample from a truncated normal distribution is given in Appendix~\ref{appendix}.

\vspace{7pt}

Fig. \ref{fig:priorty}, shows the probability distribution of the function at different values for the priority ${U_{v}}$ including 0.2, 0.4, and 0.9 and at different value for $s$. As shown in Fig.~\ref{fig:priorty a}, when $s$ = 0.4, there are overlaps between the distributions of different ${U_{v}}$. \textit{This helps in making linkability attacks difficult since ESUs with different priority levels will have requests with close priorities.} As illustrated in Fig.~\ref{fig:priorty b} and \ref{fig:priorty c}, varying the values of $s$ results in values of ${U^{(j,k)}_{v}}$ that are either more densely concentrated around ${U^{(j)}_{v}}$, as in Fig.~\ref{fig:priorty b}, or more dispersed as in Fig.~\ref{fig:priorty c}. The large values of $s$ result in dispersed priority values, and as a consequence, linking two charging requests with an ESU becomes harder. However, with large values of $s$, this will cause priorities to have a wide range of possible values, and as a consequence, the ESU's charging requests would not maintain priorities ${U^{(j,k)}_{v}}$ close to ${U^{(j)}_{v}}$, causing the charging performance to decrease. %As the values of $s$ are made smaller, linkability attacks would have better chances of succeeding since the values of ${U^{(j,k)}_{v}}$ tend to be very close to each other, consequently they could be linked to an ESU.. The opposite applies to small values of $s$. %since ${U^{k}_{v}}$ is close to ${U_{v}}$ in this case, the ESU's charging requests keep priority values proportional to  ${P_x}$.

\vspace{5pt}

Finally, Algorithm~\ref{alg:ComputeSoCandTCCvalues} gives a summary of how ESUs can compute their SoC/TCC. An ESU $v$ selects priorities $U^{(j,k)}_{v}$ for each individual request of $R^{(j,k)}_{v}$ using the random selection from the truncated normal distribution (See line 6 in Algorithm~\ref{alg:ComputeSoCandTCCvalues}). Then, it uses each priority $U^{(j,k)}_{v}$ to compute random ${S^{(j,k)}_{v}}$ and ${\mathcal{T}^{(j,k)}_{v}}$ (See line 5-13 in Algorithm~\ref{alg:ComputeSoCandTCCvalues}). Note that SoC and TCC for each request are computed randomly so that $(1)$ ${S^{(j,k)}_{v}}$ and ${\mathcal{T}^{(j,k)}_{v}}$ give ${U^{(j,k)}_{v}}$ using Eq.~\ref{1}, and $(2)$  ${\sum_{i=1}^{n}S^{(j,k)}_{v}}={S^{(j)}_{v}}$. Then, ${\mathcal{T}^{(j,k)}_{v}}$ associated with ${S^{(j,k)}_{v}}$ is computed using Eq.~\ref{1} (See line 12 in Algorithm~\ref{alg:ComputeSoCandTCCvalues}). Lastly, the last request ${S^{(j,n)}_{v}}$ is computed to maintain the equality (${\sum_{i=1}^{n}S^{(j,k)}_{v}}={S^{(j)}_{v}}$) (See line 14 in Algorithm~\ref{alg:ComputeSoCandTCCvalues}), with random priority selection of the request's priority and corresponding ${\mathcal{T}^{(j,n)}_{v}}$ (See line 16 in Algorithm~\ref{alg:ComputeSoCandTCCvalues}).

\begin{algorithm}[!t]	
\algsetup{linenosize=\small}
 \scriptsize
	\SetKwProg{Fn}{function}{}{}
	\SetKwFunction{computesSoCandTCC}{computes\_SoC\_and\_TCC}
\textbf{Input}: $U^{(j)}_{v}, n, \sigma_1, \sigma_2$\\
	${U_{v}[\ ]}$:  Array of randomly generated priorities\\ 
      ${S_{v}[\ ]}$: Array of SoC of individual charging requests \\
      ${\mathcal{T}_{v}[\ ]}$: Array of TCC of individual charging requests \\

		\For{k = 1 to n-1} {
\tcp{Random ${U^{(j,k)}_{v}}$ selection based on ${U^{(j)}_{v}}$ using truncated normal distribution}

			${U^{(j,k)}_{v}}$= Random\_Priority\_Selection(${U^{(j)}_{v}}$)

\tcp{Computing SoC of individual charging requests}

\eIf{($k$=1)}{
   $S_{v}[k]$ = rand($S_{v}$)\;
   }{
   $S_{v}[k]$ = rand($S_{v}$-$\sum_{k=1}^{n-1}$$S_{v}[k]$)\;
  }
\tcp{Computing TCC of individual charging requests}
			${\mathcal{T}^{(j,k)}_{v}}$ = $\sigma_2$/($U_{v}$-($\sigma_1$$\times$$S^{(j,k)}_{v}$)) \\	
		}

\tcp{Computing SoC and TCC of last charging request}
	 ${S^{j,n}_{v}}$ = ${S^{j}_{v}}$ - $\sum_{k=1}^{n-1}$${S^{j,k}_{v}}$\\
              ${U^{(j,n)}_{v}}$= Random\_Priority\_Selection(${U^{(j)}_{v}}$)\\
	${\mathcal{T}^{(j,n)}_{v}}$ = $\sigma_2$/($U^{(j)}_{v}$-($\sigma_1$$\times$$S^{(j,n)}_{v}$)) \\	

		\textbf{Output} (${S^{(j,k)}_{v}}[\;]$,${\mathcal{T}^{(j,k)}_{v}}[\;]$)\\
	
	\caption{Pseudocode for \textit{computing SoC and TCC} of individual charging requests}\label{alg:ComputeSoCandTCCvalues}
\end{algorithm}

\vspace{7pt}

\subsubsection{Submitting Charging Requests to the Aggregator}

After an ESU $v_i$ computes the requests SoC and TCC, it should use the anonymous tokens  previously acquired  from the CC to compose the charging requests and send them to its local aggregator. To do that, $v_i$ sends its charging request to the aggregator by sending $msg_3$

$$msg_3:= \tau^{(\ell)}_{v_i}\| 
r^{(j,k)}_{v_i}\| TS\| \sigma^{(\ell)}_{v_i}. $$

Where $msg_3$ contains a token $\tau^{(\ell)}_{v_i}=
\mathcal{M}^{(\ell)}_{v_i}  \|\sigma^{(v,\ell)}_{cc}({\mathcal{M}^{(\ell)}_{v_i}})
$, $r^{(j,k)}_{v_i}=[{S}^{(j,k)}_{v_i}\| \mathcal{T}^{(j,k)}_{v_i}]_{K^{(\ell)}_{v_{i}\leftrightarrow cc}}$ that is the SoC and TCC encrypted with the a shared secret key, time-stamp, and signature of the ESU on the entire message. Note that the signature should be done using a secret key that corresponds to a public key included in the token.

\subsection{Verification of Charging Requests and Signature Aggregation}
\label{sign_agg_ver}

When the charging requests reach the aggregator (and later the CC), they need to be verified for authenticity and integrity. As the number of ESUs in individual communities increases, the number of requests also increases, and thus more computation is needed for verifying each request signature. What makes the problem worse is that each ESU sends multiple requests. Therefore, to reduce communication overhead, once the aggregator receives ESUs' requests of its community, it should aggregate the tokens signatures as follows. 

 According to the formula in (\ref{eq: PBS sig}), the CC signature of a token $\tau^{(\ell)}_{v_i}$ takes the following form.

$$\sigma^{(v,\ell)}_{cc}(\mathcal{M}^{(\ell)}_{v_i})= \frac{\mathcal{H}_o (\mathcal{M}^{(\ell)}_{v_i} \| c)}{\mathcal{H}(c) \ + S_{cc}}     $$

Then, once the aggregator receives $\mathcal{N}$ token signature, it aggregates them into one aggregated signature $(\sigma_{agg})$ as follows.

\begin{align}
\sigma_{agg} &= \sum^{\mathcal{N}}_{i
=1}\sigma^{(v,\ell)}_{cc}(\mathcal{M}^{(\ell)}_{v_i}) \nonumber \\ & = \dfrac{\mathcal{H}_o ( \mathcal{M}^{(\ell)}_{v_1} \| c)\  + \cdots + \ \mathcal{H}_o ( \mathcal{M}^{(\ell)}_{v_{\mathcal{N}}} \| c)) }{\mathcal{H}(c)+S_{cc}} \nonumber \\
&=\dfrac{\sum_{i=1}^{\mathcal{N}}\mathcal{H}_o ( \mathcal{M}^{(\ell)}_{v_i} \| c)}{\mathcal{H}(c) \ +\  S_{cc}} 
\end{align}

Then, the aggregator sends $msg_4$ to the CC.  

$$msg_4:= \mathcal{M}^{(\ell)}_{v_i} \| r^{(j,k)}_{v_i},\cdots, \mathcal{M}^{(\ell)}_{v_{\mathcal{N}}} \| r^{(j,k)}_{v_{\mathcal{N}}}   \| \sigma_{agg} \|\sigma_{Agg}.$$

It is important to note that $msg_4$ contains all charging requests within a specific time slot. Also, it contains the aggregator' signature on the entire message $\sigma_{Agg}$.

\subsection{Computing Charging Schedules} \label{proposed_scheme}

In this phase, once the CC receives the charging requests and their aggregated signature, it verifies the aggregated signature $\sigma_{agg}$ by checking the following equality:

\begin{equation}
\label{Eq:1}
e(\mathcal{H}(c)P+P_{cc},\sigma_{agg})\stackrel{?}{=}e(P, \sum^{\mathcal{N}}_{i
=1} H_{0}(\mathcal{M}^{(\ell)}_{v_i}\| c))
\end{equation}

Also, to detect token reuse, a table of previously used tokens should be stored at the CC side. This table should include the hash of used tokens. Then, the CC checks whether the received tokens in $msg_4$ are reused. If all these verifications succeed, the CC proceeds computing the charging schedules of the requests using their SoC and TCC as follows.

Assume a community is connected to an electric bus with a maximum loading limit of $C$ and full charging request of an ESU $v$ is
($P_v$).  Our goal is to let ESUs with high priorities to charge at the present time slot, while other ESUs charging requests can be postponed to future time slots. The charging coordination problem determine whether an ESU $v$ charges in the current time slot ($y_v$) and the charging amount ($p_v$) so as to charge the ESUs with the highest priorities. Therefore, the charging coordination problem can be formulated as

\begin{eqnarray}
\begin{aligned}
 & \underset{y_{v},p_v}{\max}
 & &  \sum_{v \in \mathcal{V}} y_v U_{v}\\
 & s.t.
 & & 0 \leq p_v \leq P_v \hspace{5mm} \forall
  v \in \mathcal{V}, \\
 &&& \sum_{v \in \mathcal{V}} y_v p_v \leq C, \\
 &&& y_v \in \{0, 1\}.
     \label{2}
\end{aligned}
\end{eqnarray}

Problem (\ref{2}) is a mixed integer program (MIP) as it involves a
real variable $p_v$ and a binary variable $y_v$, which makes it
NP-complete. For a large size problem (i.e., a large community with
many ESUs), it is hard to solve (\ref{2}) in real-time.
Instead of solving the MIP in (\ref{2}), we resort to an integer
program (IP) formulation, which is less complex than (\ref{2}), and
is given by
\begin{eqnarray}
\begin{aligned}
 & \underset{y_{v} \in \{0, 1\}}{\max}
 & &  \sum_{v \in \mathcal{V}} y_v U_{v}\\
 & s.t.
 & & \sum_{v \in \mathcal{V}} y_v P_v \leq C.
     \label{3}
\end{aligned}
\end{eqnarray}

According to (\ref{3}), if an ESU is scheduled to charge during
the current time slot, it receives its full charging request
($P_v$) in the current (single) time slot. The scheduling problem in (\ref{3}) can be mapped to an optimization problem referred to as the knapsack problem \cite{R}. In the knapsack problem, there is a knapsack with limited capacity and a set of items each with a given value (priority) and weight. The goal is to choose a subset of items to be packed in the knapsack, such that the total value is maximized while the knapsack capacity limitation is respected. The charging coordination problem can be mapped to a knapsack problem as follows. First, the ESUs are mapped to the items. Then, the ESU priority $U_v$ is
equivalent to the item value. And, the ESU charging demand $P_v$ is
equivalent to the item weight. Finally, the charging capacity limitation $C$ is equivalent to the knapsack capacity. A greedy algorithm for solving the knapsack problem in polynomial time complexity can be used to schedule ESU charging during a given time slot \cite{R}. Hence, the charging coordination mechanism can be described using Algorithm \ref{alg1}, which is executed by the CC.

Once the CC finalizes the charging coordination, it prepares the charging schedules $\{Sc^{(j,k)}_{v_1}, Sc^{(j,k)}_{v_2},\cdots\}$, where $(Sc^{(j,k)}_v=P K_{v}^{(\ell)},[y_v, p_v]_{K^{(\ell)}_{v \leftrightarrow cc}}$) should be encrypted using the one-time key ($K^{(\ell)}_{v\leftrightarrow cc}$) sent by the ESU. Then, the CC sends the following message $msg_5$ to the aggregator

$$msg_5=\{Sc^{(j,k)}_{v_1}, Sc^{(j,k)}_{v_2},\cdots\}\| \sigma_{CC}.$$ 

Finally, once the aggregator received $msg_5$, it broadcasts the message to the community. Each ESU knows its charging schedule that corresponds to the public key included in the charging schedule. Then, it can determine to charge by decrypting the schedule using the shared secret key.

\begin{algorithm}[!t]%[PEV Charging Coordination Mechanism]
\caption{ESU Charging Coordination Mechanism \label{alg1}     }  %\label{alg1}
%\begin{algorithmbis}
\begin{algorithmic} [t] \STATE \textbf{Input}: $\mathcal{V}$, $U_v$ and $P_v$ $\forall v \in \mathcal{V}$; \STATE \textbf{Initialization}: $y_v = 0$ $\forall v \in \mathcal{V}$, $\mathcal{A} =
\{\}$, $C_R \leftarrow C$; \STATE Sort all ESUs in
$\mathcal{V}$ such that $\frac{U_1}{P_1} \geq \frac{U_2}{P_2} \ldots
\geq \frac{U_V}{P_V}$ and store the result in $\mathcal{A}$; \FOR
{$v \in \mathcal{A}$} \IF {$P_v \leq C_R$} \STATE $y_v = 1$; \STATE
$p_v = P_v$; \STATE $C_R = C_R - P_V$; \STATE $\mathcal{A} = \mathcal{A} - \setminus \{v\}$; \ENDIF \ENDFOR \STATE $L = \underset{\mathcal{A}} {\mathrm{argmax}}$ $U_v$; \STATE $y_L = 1$;
\STATE $p_L = C_R$; \STATE \textbf{Output}: $X$ and $P$.
\end{algorithmic}
%\end{algorithmbis}
\vspace{-0.1cm}
\end{algorithm}

\section{The DCC Scheme}
\label{sec: decentralized}

In this section, we first give an overview to our scheme and then we discuss  charging requests submission and aggregation, and charging schedules computation.

\subsection{Overview}

\begin{figure}[!t]
  \begin{center}
\setlength{\abovecaptionskip}{-0.2cm}
\includegraphics[width=.8\linewidth]{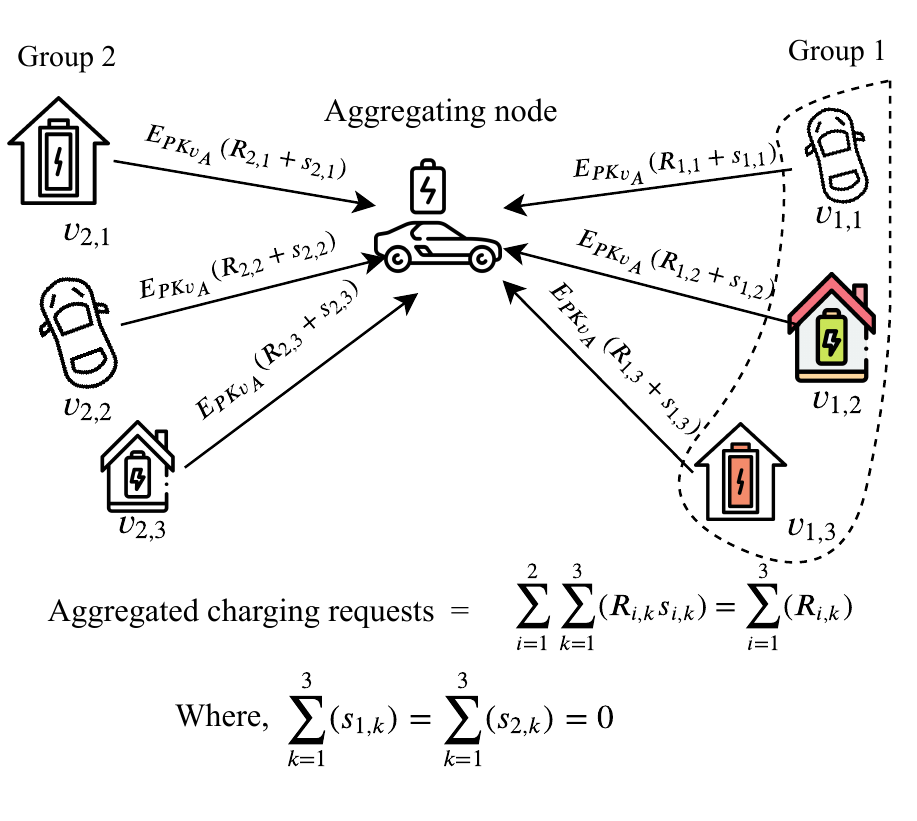}

\caption{Illustrative example of the data masking technique. By
aggregating all requests, all masks are nullified and the total charging demand for each priority level can be computed.} \label{fig:decentralized exhanged}
\end{center}\end{figure}

The DCC scheme runs in a fully distributed way where several ESUs should run the scheme to collect the total amount of power and in the same time coordinate charging demands. However, the challenge is how a group of ESUs can collect individual requests, aggregate them, disseminate the aggregated demands, and compute the charging schedules without support from infrastructure and with protection against collusion attacks. In other words, although the ESUs are not trusted, they should run the scheme securely and without leaking sensitive information. In each time slot, a number of ESUs are selected to act as aggregating ESUs. Each ESU selects a set of ESUs called \textit{proxies} and share a mask with each one. Each ESU then adds the mask to its charging request, encrypts it using homomorphic encryption, and sends it to the aggregating
node as shown in Fig.~\ref{fig:decentralized exhanged}. After that, the aggregating nodes decrypt the ciphertext of the aggregated message of the ESUs requests and broadcast it to the community ESUs but without being able to access each ESU demand to preserve privacy. Then, charging schedules are calculated locally by each ESU to know whether it can charge in the time slot and the amount of power it can charge. Note that, unlike the CCC scheme, in the DCC, each ESU sends only one charging request in each time slot since no one have access to the charging demand of a specific ESU and only the total aggregated demand is known.

\begin{figure}[!t]
  \begin{center}
\setlength{\abovecaptionskip}{-0.2cm}
\vspace{1mm}
\includegraphics[width=0.92\linewidth]{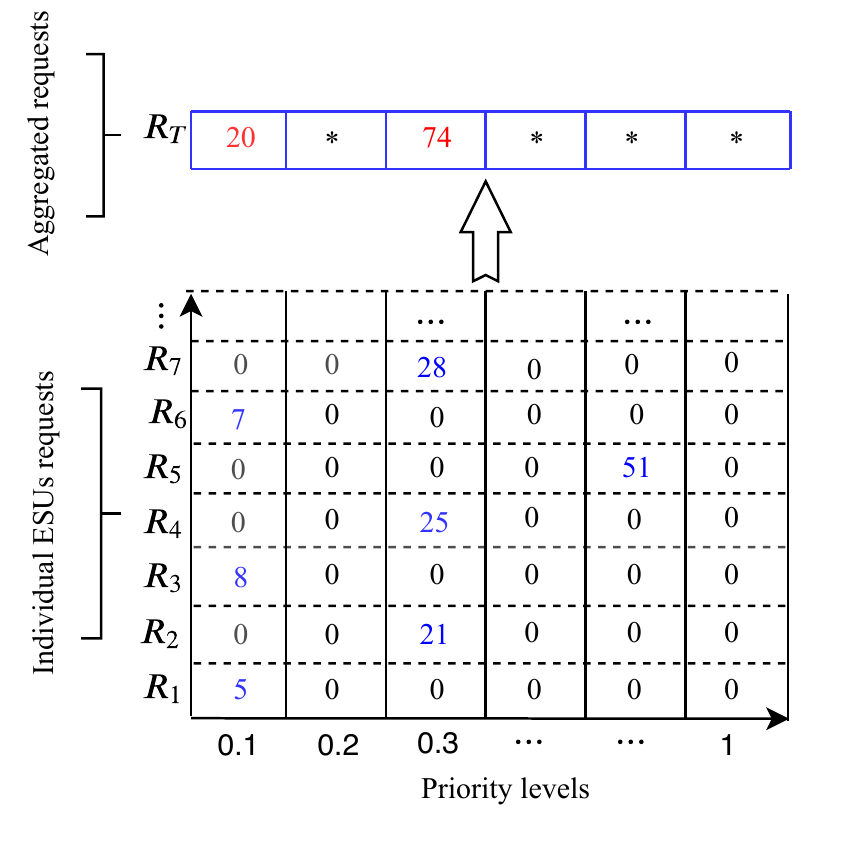}
\caption{Charging requests structure and the data aggregation technique. The sum of the bits at i-th column represents the total charging demand needed by ESUs at a priority level $L_i$. For simplicity, the charging requests are represented in decimal rather than binary.  } 

\label{fig:ChargingPrioritysets}
\end{center}\end{figure}

To illustrate the idea, each request is represented as an field element that can be encrypted using Paillier Cryptosystem. Each charging request message is divided into priority levels (e.g., ${L_1}$ $\in [0,0.1), L_{2} \in [0.1,0.2),\cdots,$ and $L_{10} \in [0.9,1])$, and each priority level has a set of associated bits that are used to report the amount of charging power the ESU needs if it has the priority level of the set. For instance, as in Fig.~\ref{fig:ChargingPrioritysets}, consider for simplicity a field element of size 1000 bits, and 10 priority levels $L_1$ to $L_{10}$, then a field element can be divided into 10 \textit{set of bits} each of them is of size 100 bits. Also, consider a request from an ESU $v_1$ denoted as $R_1$ = 5 KW, and the priority of an ESU calculated using Eq. \ref{1} lies in level $L_1$,  the ESU will use the right most 100 bits to write its charging needs and all other bits should be set as zero. Note that since all messages should be aggregated and to avoid arithmetic overflow, each priority level should be assigned sufficient bits to avoid adding a carry for the next priority level. By this way, the aggregated message gives the total charging demand for each priority level (See $R_T$ in Fig.~\ref{fig:ChargingPrioritysets}).

\subsection{Charging Requests Submission and Aggregation}
In this phase, ESUs submits their charging messages to the aggregating node.

First, each ESU $v_i$ should generate a public/private key pair $Y_{v_i}$ and $x_{v_i}$, where the private key $x_{v_i}$ $\overset{R}{\leftarrow}$ $\mathbb{Z}^*_q$, the public key is $Y_{v_i}$ = $x_{v_i}{P}$, ${q}$ is a large prime number, and ${P}$ is the generator of a cyclic additive group ${G_1}$. Also, the aggregating node should generate public key $(N_{v_A}, g_{v_A})$ and corresponding private key $\delta_{v_A}$ for homomorphic encryption scheme, and broadcast the public key. Then, each ESU $v_i$ chooses a shared secret mask $s_{v_i}$ with a group of ESUs called \textit{proxies} such that the ESU $v_i$ shares a mask $s_{v_{i,j}}$ with proxy ESU $v_{j}$ for $1\leq j\leq k$ where $k$ is the number of proxy ESUs such that 
$$s_{v_i}=\sum_{j=1}^{k}s_{v_{i,j}}$$

Then, to report a charging message $R^{j}_{v_i}$ at a time slot $j$, each ESU $v_i$ uses the public key of the aggregating node's homomorphic encryption ($N_{v_{A}}, g_{v_{A}}$) and a random number $r \in Z_{n}^{*}$  where $r$ is known to all ESUs (e.g., $r=H(date\|SN)$ where $H$ is a hash function, $date$ is the current date, and $SN$ is a counter for the time slot of the day) to compute $\mathcal{C}^{j}_{v_i}$ which is the encryption of $R^{j}_{v_i}$ as follows.

$$
\mathcal{C}^{j}_{v_i}=g_{v_A}^{R^{j}_{v_i}} r^{({N_{v_A}+s_{v_i}})} \bmod N_{{v_A}}^{2}.
$$

After that, the ESU signs $\mathcal{C}^{j}_{v_i}$ after appending a time stamp by computing $\sigma_{v_i}=x_{v_i} H\left(\mathcal{C}^{j}_{v_i} \| TS\right)$. The ESU sends to the aggregating node the following packet $\mathcal{C}^{j}_{v_i}\|T S\| \sigma_{v_i}$.

After the $n$ ESUs in the neighborhood report their messages to the aggregating node, it  verifies the charging message's signatures with less overhead (fewer number of pairing operations), using a batch verification technique~\cite{e22}. In this technique, instead of verifying $n$ individual signatures, the signatures can be batched and one verification process is executed. The signatures are valid if $e\left(P, \sum_{i=1}^{n} \sigma_{v_i}\right)=\prod_{i=1}^{n} e\left(Y_{v_i}, H\left(\mathcal{C}^{j}_{v_i} \| TS\right)\right)$. The proof
for this is as follows:

$$
\begin{aligned} e\left(P, \sum_{i=1}^{n} \sigma_{v_i}\right) &=e\left(P, \sum_{i=1}^{n} x_{v_i} H\left(\mathcal{C}^{j}_{v_i} \| T S\right)\right).\\ &=\prod_{i=1}^{n} e\left(P, x_{v_i} H\left(\mathcal{C}^{j}_{v_i} \| T S\right)\right) \\ &=\prod_{i=1}^{n} e\left(Y_{v_i}, H\left(\mathcal{C}^{j}_{v_i}\| T S\right)\right) \end{aligned}
$$

Finally, if the signature verification succeeds, the aggregating node should aggregate the ciphertexts as follows:

$$
\begin{aligned} \mathcal{C}^{j}_{T} &=\mathcal{C}^{j}_{v_1} \times \mathcal{C}^{j}_{v_2} \times \ldots \times \mathcal{C}^{j}_{v_n} \\ &=g_{v_A}^{R^{j}_{v_1}} r^{\left(N_{v_A}+s_{i}\right)} \bmod N_{{v_A}}^{2} \times \ldots \times g_{v_A}^{R^{j}_{v_n}} r^{\left(N_{v_A}+s_{n}\right)} \bmod N_{{v_A}}^{2} \\ &=g_{v_A}^{\sum_{i=1}^{n} R^{j}_{v_i}} \times r^{\sum_{i=1}^{n} s_{v_i}} \times {r^{(n)^{N}}}\\ &=g_{v_A}^{\sum_{i=1}^{n} R^{j}_{v_i}} \times r^{0} \times {r^{(n)^{N}}}=g_{v_A}^{\sum_{i=1}^{n} R^{j}_{v_i}} \times {r^{(n)^{N}}}\end{aligned}
$$

 Then, the aggregating node should use its private key $\left(\delta_{v_A}\right)$ to decrypt $\mathcal{C}^{j}_{T}$ and obtain the total aggregated charging demand  $R^{j}_T=\sum_{i=1}^{n} R^{j}_{i}$ at a time slot $j$ by computing $\frac{L\left(\mathcal{C}_{T}^{\delta_{v_A}} \bmod N_{v_A}^{2}\right)}{L\left(g_{v_A}^{\delta_{v_A}} \bmod N_{v_A}^{2}\right)} \bmod N_{v_A}$. By aggregating all the messages, the set of bits that correspond to a priority level $L_i$, gives the total charging demand needed by all the ESUs in the community that have a priority level $L_i$. %Note that the maximum charging capacity $C$ can be determined by the same way of calculating ($R_T$).

It is important to note that the purpose of adding a mask by an ESU $v_i$ and removing it by other proxy ESUs messages is to:
\begin{enumerate}

\item Prevent ${v_A}$ from knowing the demand of one ESU because given  $\mathcal{C}^{j}_{v_i}=g^{R^{j}_{v_i}}r^{{N}+{s_i}}$, ${v_A}$ cannot use its private key to decrypt the message because the exponent of $r$ should be $N$, but given $\mathcal{C}^{j}_{T}=g^{\sum_{i=1}^{n}{R^{j}_{v_i}}} ({r^{n}}){^N}$, ${v_A}$ can decrypt this message using its private key. Note that $r^n$ or $r$ does not make a difference in the decryption because both are random numbers. 
\item Protect against collusion attack since to get the charging demand of an ESU $v_i$, ${v_A}$ has to collude with all the proxy ESUs to obtain ${s_{v_{i,j}}}$ for ${j=\{1,2,...,k\}}$ to compute ${s_{v_i}}$ and then compute ${R^{j}_{v_i}}$ as follows:
$$r^{-s_{v_i}} \mathcal{C}^{j}_{v_i}={g^{R^{j}_{v_i}}r^{N+{s_{v_i}}}}{r^{-s_{v_i}}}={g^{R^{j}_{v_i}}}r^N$$
\end{enumerate}

Then, $v_A$ uses its private key to decrypt $g^{R^{j}_{v_i}}r^N$ to obtain $R^{j}_{v_i}$. 

 For simplicity, it is assumed that there is only one aggregating node, but the scheme can easily be extended to have multiple aggregating nodes. Therefore, if one aggregating node is not trusted in reporting the total demand correctly, aggregation can be performed by more than one aggregating node and the total demand can be broadcasted by the aggregating nodes. Also, the aggregating nodes should continuously change to distribute the decryption overhead on other community nodes.

\subsection{Charging Schedules Computation}
After ${v_A}$ computes the aggregated charging message ($ R^{j}_T\leq C^{j}$), the aggregating node broadcasts it to all the ESUs. Then, if $ R^{j}_T\leq C^{j}$, where $C^{j}$ is the maximum charging capacity, then all requests are granted to charge since there is enough energy to serve the demands of all the ESUs. If $R^{j}_T > C^{j}$, then ESUs with highest priority should charge without exceeding the maximum capacity $C^{j}$. This is done as follows. Assume $L_i$ is the priority level at which ESUs can charge. Each ESU compares $C^{j}$ to the total charging demand of priority levels, from the highest level to the lowest level, until it finds the first level $L_{i}$ at which the total accumulated demand is greater than or equal to the maximum charging capacity. In other words, $L_i$ is the lowest priority level that guarantees the condition $\sum_{l=L_i}^{L_{max}}{R_{T}^{(j,l)}}$ $\leq C^{j}$, where $L_{max}$ is the maximum priority level, and ${R_{T}^{(j,l)}}$ is the total charging demand of a priority level at a time slot $j$. If the total charge capacity is equal to the total charging demand of these priority levels, then all the ESUs that have priority level greater than or equal to $L_{i}$ can charge and the other ESUs do not charge in this time. If the power demand of all the sets from the highest priority to priority level $L_{i}$ is greater than the total charging capacity, then all the ESUs that have priority greater than $L_{i}$ should charge and for full utilization of the available charging power capacity. The remaining power is charged by the ESUs of priority level $L_{i}$. The power $E_{v_i}$ that an ESU $v_i$ of the priority level $L_{i}$ should charge is given by:

\begin{equation}
\label{eq:deltafrac}
E_{v_i}=\Delta \times \frac{R^{j}_{v_i}}{R_T^{(j,L_{i})}},
\end{equation}

where $R_T^{(j,L_{i})}$ is the demand energy by the ESUs with priority level $L_{i}$, and $\Delta$ is given by 

\begin{equation}
\label{eq:delta}
\Delta = \ {C^{j} - \sum_{l=L_i}^{L_{max}}{R_T^{(j,l)}}}.
\end{equation}

To illustrate the idea of charging schedules computation, we present a numerical example using ten ESUs and a community capacity $C$ of 300KW. Table \ref{table 1} gives the power demands and priorities of each ESU that are selected arbitrarily. Table \ref{table 2} gives the corresponding charging requests of all ESUs. Once the aggregating node ${v_A}$ computes the total charging message (As shown in the last row of~Table \ref{table 2}), the total demand is broadcasted to all ESUs in the community. After receiving this message, each ESU finds the priority level $L_{i}$, and determine whether they can charge or cannot charge, to reduce their charging demands by using Eq.~\ref{eq:deltafrac}. In this example, $L_i$ is level 4, since $\sum_{l=4}^{10} {R^{(l)}_T}$ = 210KW, and including level 3 would result in exceeding the capacity since $\sum_{l=3}^{10} {R^{(l)}_T}$ = 310KW. By using the formula in~\ref{eq:delta}, $\Delta$ = 300KW - 210KW = 90KW. Because priority level 3 is the first level exceeding the capacity, all ESUs with at least level 4 can charge. Additionally, those at level 3 i.e., ESUs $2$ and $10$ compute the amount of energy they can charge according to use formula in~\ref{eq:deltafrac} as follows, 

\begin{equation}
\label{eq:deltafrac1}
E_{v_2}=90 \times \frac{30}{310}= 27,   E_{v_{10}}=90 \times \frac{70}{310}=63,\nonumber
\end{equation}

and those with level 2 or below, should not charge in the current time slot and need to submit new charging requests in the next time slot.

Finally, Table \ref{table 3} gives the charging schedules, where the second column gives the  initial charging request, and the third column gives the charging schedule including the amount of power each ESU can charge.

\begin{table}[!t]
\caption{Demand and final charging schedules.}
	\label{table 1}
\begin{center}
\scalebox{0.99}{
\begin{tabular}{|c|c|c|c|c|}
 \hline

 & ESU $v_i$&\makecell{Power demand\\(KW) ($R_i$)}&Priority ($U$)&Level ($L$)\\
 \hline
   & 1    &10&   0.333&4\\
 &  2  & 30 & 0.250&3\\
  &3& 50&  1&10\\
  &4 &60&  0.166&2\\
 &  5  & 90&0.333&4\\
 & 6  &20 &0.143&2\\
&7& 5&  0.143&2\\
  &8 & 40&  0.500&6\\
 &  9  & 20&1&10\\
 & 10 & 70&0.200&3\\
\hline 
 \textbf{Total}& & \textbf{395}&&\\
 \hline
\end{tabular}
}
\end{center}
\end{table}

\begin{table}[!t]
\caption{Charging packets and priority levels.}
\label{table 2}
\begin{center}
\scalebox{0.93}{
\begin{tabular}{|c|c|c|c|c|c|c|c|c|c|c| }
 \hline
 ESU&$L_{10}$&$L_9$&$L_8$&$L_7$&$L_6$&$L_5$&$L_4$&$L_3$&$L_2$&$L_1$\\
 \hline
 1 & 0   &0&   0&0   &0&   0&10   &0&   0&0   \\
 \hline
 2 & 0   &0&   0&0   &0&   0&0   &30&   0&0   \\
\hline
 3 & 50   &0&   0&0   &0&   0&0   &0&   0&0   \\
\hline
 4 & 0   &0&   0&0   &0&   0&0   &0&   60&0   \\
\hline
 5 & 0   &0&   0&0   &0&   0&90   &0&   0&0   \\
\hline
 6 & 0   &0&   0&0   &0&   0&0   &0&   20&0   \\
\hline
 7 & 0   &0&   0&0   &0&   0&0   &0&   5&0   \\
\hline
  8 & 0   &0&   0&0   &40&   0&0   &0&   0&0   \\
\hline
 9& 20   &0&   0&0   &0&   0&0   &0&   0&0   \\
\hline
 10 & 0   &0&   0&0   &0&   0&0   &70&   0&0   \\
 \hline
 \hline
$R_{T}$& 70   &0&   0&0   &40&   100&  0 &100&   85&0  \\
 \hline
\end{tabular}}
\end{center}
\end{table}

\begin{table}[!t]
\caption{Initial and final charging.}\label{table 3}
\begin{center}
\scalebox{0.92}{
\begin{tabular}{|c|c|c|}
 \hline
 
 ESU $v_i$  & \makecell{Original\\ demand} $({R_i})$ & Schedule  $(E_{i})$\\
 \hline
 1 & 10   &10\\
 \hline
 2 & 30    &\textbf{27}\\
\hline
 3 & 50   &50\\
\hline
 4 & 60    &0\\
\hline
 5 & 90    &90\\
\hline
 6 & 20   &0\\
\hline
 7 & 5   &0\\
\hline
  8 & 40   &40\\
\hline
 9& 20   &20\\
\hline
 10 &70  &\textbf{63}\\
\hline
Total&395&300\\
\hline
\end{tabular}}
\end{center}
\end{table}

\section{Security and Privacy Analysis}
\label{Sec: security and privacy}
In this section, we discuss and analyze the security and privacy of our schemes. 
\subsection{CCC scheme}

\subsubsection{Resistance to Linkabilty Attacks}

 By submitting multiple requests with random SoC and TCC instead of only one request, it is hard for the CC to link requests sent from the same ESU using SoC and TCC. However, since the priorities of requests of an ESU are proportional to its priority, the CC could use this information to attempt linkability attacks on charging requests. However, due to the probabilistic nature of priorities of the charging requests, the priorities of the charging requests of different ESUs may overlap which can confuse the CC and make the linkability difficult. 

In order to assess our scheme under various linkability attack scenarios, different linkablity attacks are considered in our evaluation; $(i)$ Using both SoC and TCC; and $(ii)$ Using charging requests priorities.

We have used Matlab to evaluate success probability of different linkability attacks. We set the number of ESUs to 80, and the community charging capacity was set at 1000kW. We set the ESUs battery capacity to 100 kW, and the SoC of each ESU battery is a random number uniformly distributed in $[0, 1]$. The TCC range was set to a random number in $\{1,\cdots, 48\}$ and a total of 30 time slots are simulated. The values of $\alpha_1$ and $\alpha_2$ were selected as 0.9 and 0.1 respectively.

\begin{figure*}[!t]
	\setlength{\abovecaptionskip}{0.1cm}
	\setlength{\belowcaptionskip}{-0.7cm}
	\centering
	\subfloat[Using both SoC and TTC of charging requests.\label{7a}]
    {\includegraphics[width=0.34\linewidth]{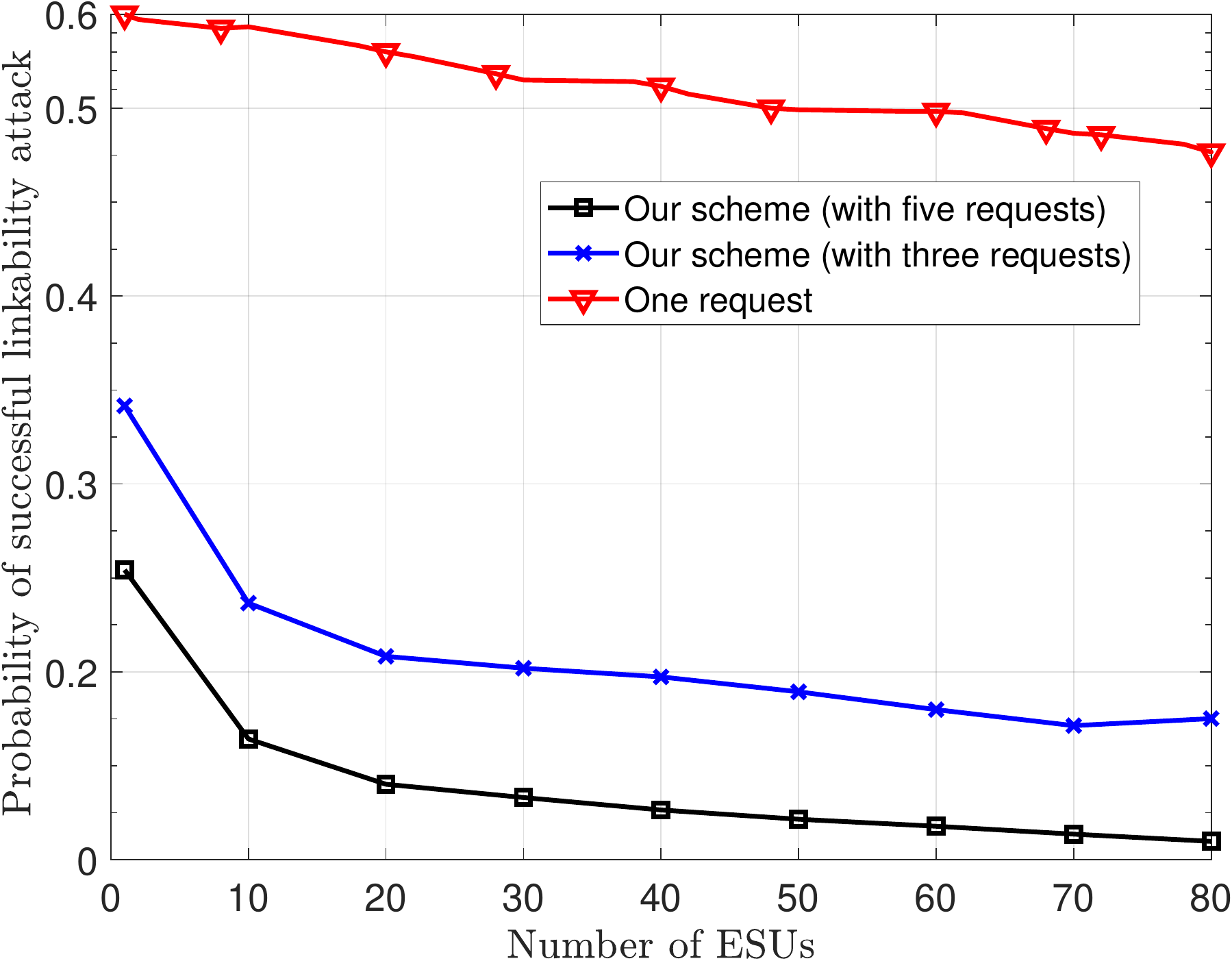}}\hspace{35px}
    \subfloat[Using charging request's priority.\label{7b}]
	{\includegraphics[width=0.34\linewidth]{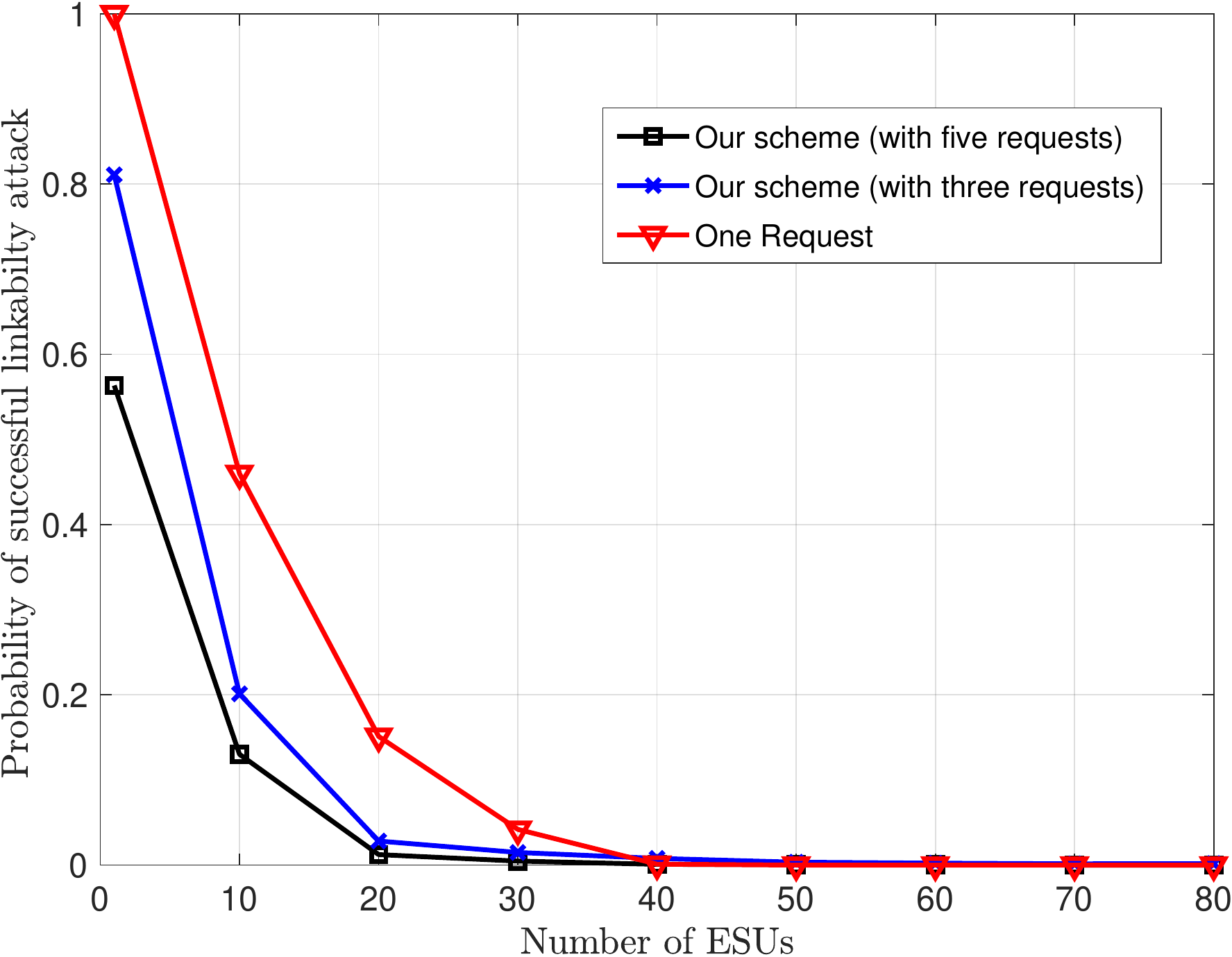}}
 	\caption{Evaluation of linkability attacks using different cases.}
	\label{fig:computation_new}
	\vspace{10px}
\end{figure*}

The simulations results depicted in Fig.~\ref{7a} presents the probability of successful linkability attack using SoC and TTC. It can be clearly seen that our scheme reduces the ability of CC to launch a successful linkability attack using SoC and TTC at different time slots. Also, as the number of requests per ESUs increases, as the probability of successfully linking charging requests decreases. This is because as the number of requests per ESU increases, as the overlap between charging requests increase making linkability attack less successful.

 %The likelihood of successful linkability attack is very close to zero with 20 ESUs in our scheme, as opposed to being close to one even when the number of ESUs is 80 without our scheme.

In case of using priorities of ESUs to launch linkability attacks, Fig.~\ref{7b} shows that low success probability compared to the previous case, i.e., using SoC and TCC. This is because of using of short-range values of priorities and the use of the non linear priority function ( Eq.~\ref{1}) that transforms the SoC and TCC values into priority values that could be close to each other. Also, in case of an increasing number of requests per ESU, results show similar performance since our scheme ensures priority of several requests overlap, making it very hard to the CC to use the priority to link charging requests of ESUs at different time slots.

\subsubsection{Resistance to Other Attacks}

\begin{itemize}
 
 \item \textit{Resistance to collusion attacks.} By using PBS during the acquisition of anonymous tokens, each ESU can anonymously authenticate its charging requests sent to the aggregator without the need to reveal its real identity.  If the CC and the aggregator collude, they cannot infer the SoC/TCC of an ESU because the CC signature on the tokens is not linkable. Moreover, by using a one-time generated identity that is not linkable to an ESU, the privacy of the ESUs is preserved.
 
\item \textit{Data authentication.} All the messages in our scheme are authenticated to ensure that legitimate ESUs send those messages. This authentication is done using signatures.  Each ESU obtains anonymous tokens signed by the CC. The CC verifies the signature once a token is used along with a charging request. This signature on the anonymous token guarantees that the request is sent by a legitimate ESU member of a community, as only legitimate ESUs in this community could obtain valid tokens from the CC because the ESU should sign to obtain tokens.

 \item \textit{Resistance to replay attacks and token reuse.}  To prevent replay attacks, all the messages in our scheme have a signed fresh timestamp. In addition, since the CC keeps a record of previously used tokens by storing their hash values; if an ESU attempts to reuse a token, the CC can detect and discard this request. 

 \item \textit{Confidentiality of charging schedules.}  The SoC and TCC are encrypted with a secret key that is only known to the CC and the ESU that sends the charging request. Also, the encryption key is encrypted by the CC's public key and the ciphertext is sent to the CC. This ensures that only the CC can decrypt the ciphertext and use the symmetric encryption key when sending the charging schedule back to the ESU. By this way, only ESUs can know their charging schedules.

\end{itemize}

\subsection{DCC scheme}
\begin{itemize}

  \item \textit{Privacy preservation.} In DCC, any ESU cannot know charging report messages of other ESUs. This is done by aggregation and secret masks addition. By using aggregation, only the total charging report can be known. Also, due to the aggregation technique, the aggregating node (s) cannot link charging requests to a specific ESU owner in consecutive time slots. By this way, linkabilty in the DCC is mitigated.

 \item \textit{Resistance to collusion attacks.} By using secret masks to mask the ESUs' messages, the aggregating node that knows the private key of the homomorphic encryption scheme has to collude with a number of ESUs (proxy ESUs) to decrypt the message, using the mask shared with the proxy ESUs. With using enough number of proxy ESUs, the attack can be infeasable because the head ESU has to collude with a large number of proxy ESUs. To evaluate the protection against collusion, we define a probability formula. If ${m}$  ESUs are malicious nodes colluding with the aggregating node from a total of ${n}$ ESUs in the community, with ${m}$ $>$ $\delta$, and $\delta$ is the number of proxy ESUs, the probability of ${v_A}$ colluding with all $\delta$ ESUs out of a total of ${n}$ ESUs,  follows a hypergeometric distribution. The probability distribution function (PDF) of the hypergeometric probability distribution is given by:

\begin{centering}
\[
    pdf(x|n,m,\lambda) = \frac{\binom{m}{x} \binom{n-m}{\lambda-x}}{\binom{n}{\lambda}}
\]

where $x$ is the number of malicious ESUs included in the selection of $\lambda$ ESUs out of ${n}$ ESUs. This probability distribution corresponds to the number of successful selections of $\lambda$ proxies among $m$ malicious ESUs.
\end{centering}

The probability of the ${v_A}$ colluding with all ${\lambda}$ proxy nodes is plotted in Fig.~\ref{fig:HyperGeoCollusion} based on various values of ${m}$ and $\lambda$, where the number of ESUs in the network is ${n}$ = 300. From Fig.~\ref{fig:HyperGeoCollusion}, it can be clearly seen that when an ESU selects $\lambda$ = 4  proxies, roughly 1\% out of the total number of ESUs in the community, and the number of malicious ESUs is ${m}$ = 100, or 30\% of the total ${n}$ ESUs, the probability of revealing a charging demand of an ESU by ${v_A}$ is 0.07,  the probability of revealing a charging demand of an ESU drops to 0.01 for the case of selecting 8 proxy ESUs. \textit {As the number of proxies increases, it becomes more difficult for ${v_A}$ to recover the charging demands of the ESUs}. Note that if the number of proxies selected is $\lambda$ = 16, the probability of recovering a charging demand by ${v_A}$ becomes around zero. This is under the assumption that there are already 100 malicious nodes colluding with ${v_A}$, which in reality is a very large number. If the number of malicious ESUs is increased to 200, or 66\% of the total ESUs in the community, then the probability of revealing a charging request data by the  ${v_A}$ becomes one. \textit{If further protection is needed, then the number of proxy ESUs should be increased}.

    \begin{figure}[t!]
  \begin{center}
\setlength{\abovecaptionskip}{-0.2cm}

\includegraphics[scale=0.4]{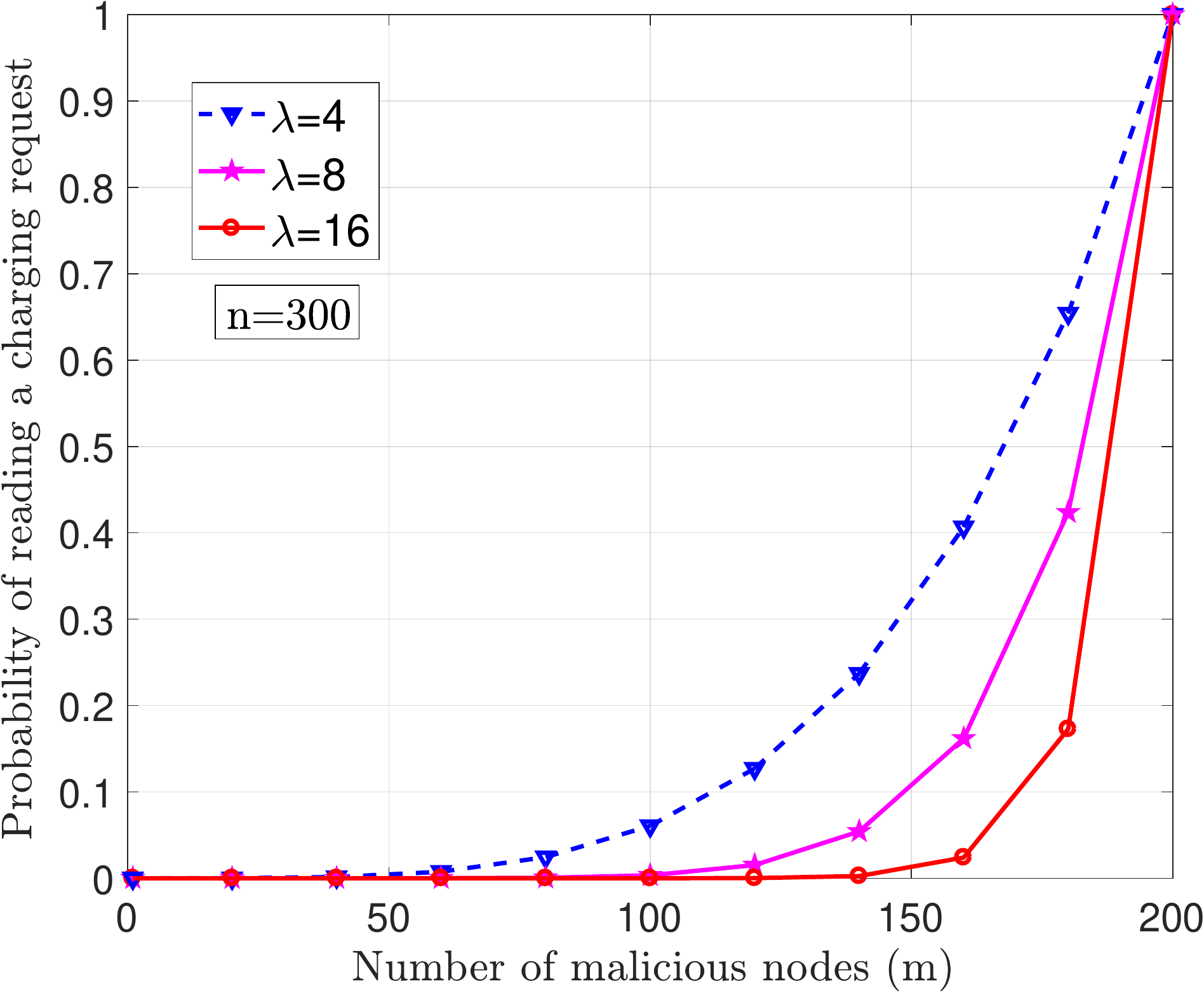}
\vspace{5mm}
\caption{Probability of revealing a charging request data by colluding with ${\lambda}$ proxies.} \label{fig:HyperGeoCollusion}
\end{center}

\end{figure}

\end{itemize}

\section{Performance Evaulations}
\label{Sec:  charging evaulations}
In this section, we evaluate the charging coordination schemes followed by the communication and computaion overheads.

\subsection{Charging Coordination}

\subsubsection{Metrics and Baselines}
In both schemes, we consider \textit{charging index} as the performance metric for charging coordination. Charging index is defined as the number of ESUs with charging requests that expire without fully charging. Lower values of  this metric indicate higher ESU charging energy and hence efficient utilization of the energy resources and smarter scheduling of ESUs charging. We compare our charging coordination schemes with first come first serve (FCFS) benchmark. In this approach, the ESU that requests charging first gets charged first.

For both schemes, the capacity ${C}$ is assumed to be 1,000kW, with the priority of each charging request per ESU is computed using Eq.~\ref{1} with TCC values are selected randomly from 1 to 48, and SoC values are selected randomly from 1 to 100kW. For the DCC, ten different levels of priorities are used, where the level $L_1$ corresponds to [0, 0.1), level $L_2$ corresponds to [0.1, 0.2), and so on until $L_10$, which corresponds to the range of priorities [0, 1]. The number of ESUs was varied with increments of one in each run, ranging from 1 to 80, and 30 time slots were considered. For the CCC and DCC, 100 runs were performed over a period of 30 time slots and the average is presented.

\begin{comment}
To evaluate the charging coordination for the DCC, simulation set up were conducted as follows. The capacity ${C}$ is assumed to be 1,000kW, with the priority of each charging request per ESU was computed using Eq.~\ref{1} with TCC values are selected randomly from 1 to 48, and SoC values are selected randomly from 1 to 100kW. In addition ten different levels of priorities are used, where level 1 corresponds to [0, 0.1], level 2 corresponds to [0.1, 0.2], and so on until level 10, which corresponds to the range of priorities [0, 1]. The number of ESUs was varied with increments of one in each run, ranging from 1 to 45, and 30 time slots were considered. For each number of ESU, 100 runs were performed and the average is presented in the given results.
\end{comment}

\subsubsection{CCC Scheme}

\begin{comment}

To evaluate the charging coordination in the centralized scheme, the simulation parameters were set as follows. The community capacity ${C}$ was set to 1000kW, and the number of ESUs are ranged from 1 to 45 with increments of one in every run. The maximum energy demand  per ESU is set to 100kW, with the SoC and TCC of the ESUs assigned randomly using  uniform distribution, where SoC $\in$ [0,1] and TCC $\in$ $\{1,2,\cdots,48\}$. Each run was conducted over a period of 30 time slots. 

\end{comment}

\begin{figure*}[!t]
	\setlength{\abovecaptionskip}{0.1cm}
	\setlength{\belowcaptionskip}{-0.7cm}
	\centering
	\subfloat[CCC scheme.\label{6a}]
    {\includegraphics[width=0.35\linewidth]{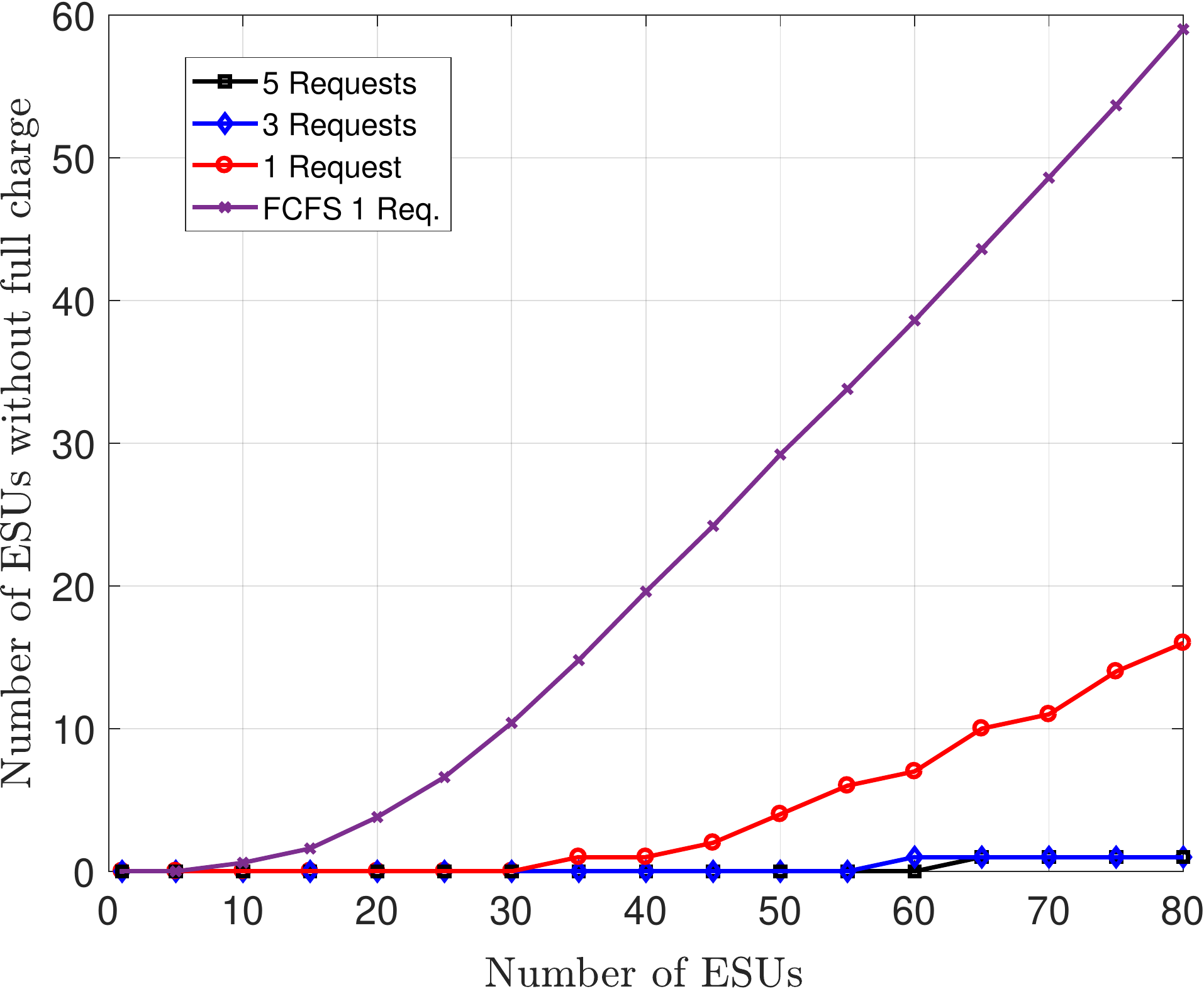}}\hspace{25pt}
    \subfloat[DCC scheme.\label{6b}]
	{\includegraphics[width=0.35\linewidth]{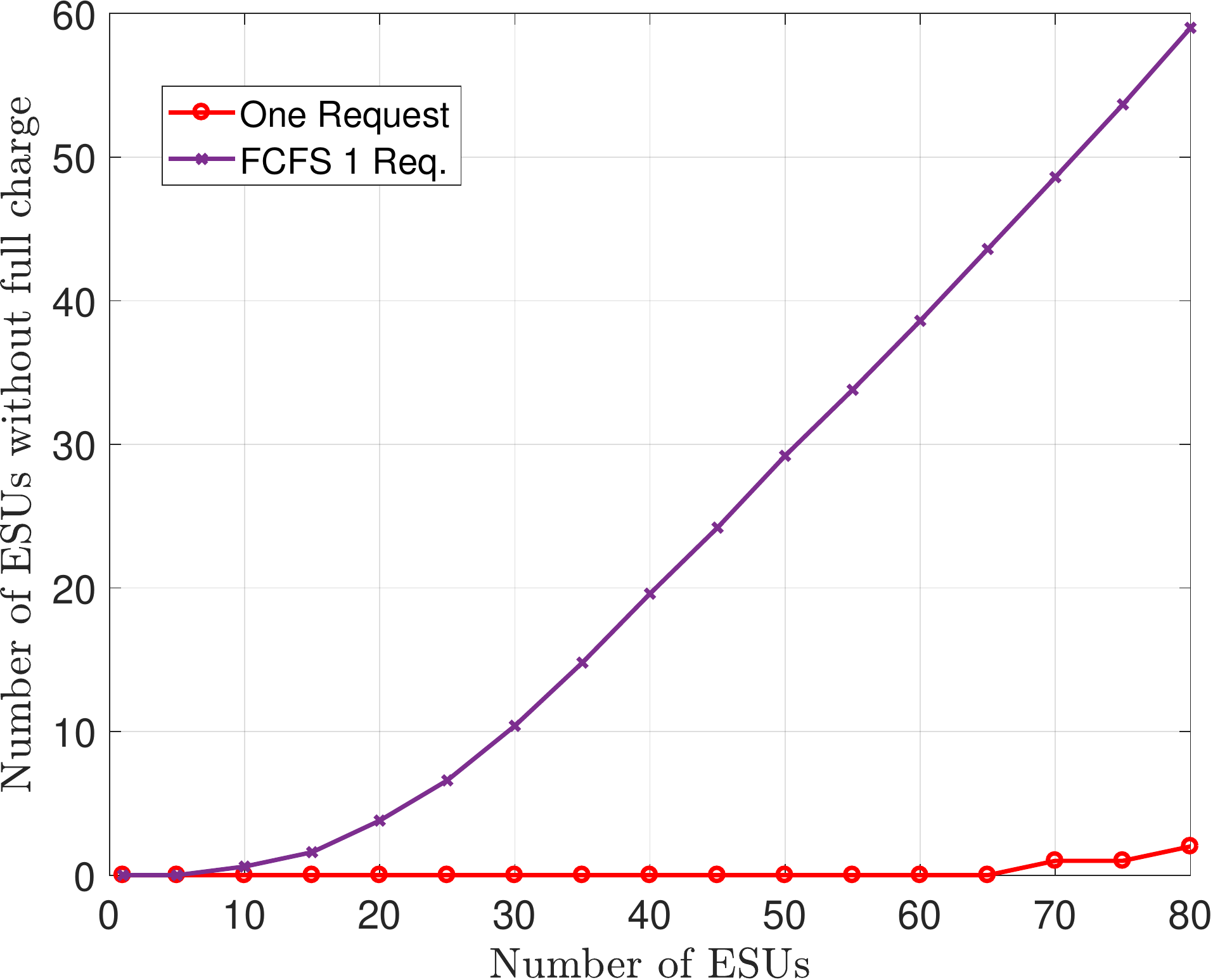}}
   
	\caption{Charging coordination evaluation.}
	\label{fig:computation_new}
	\vspace{6pt}
\end{figure*}

Fig.~\ref{6a} gives the number of ESUs that do not fully charge before TCC expires versus the number of ESUs. It can be seen that as the number of charging requests increases, our scheme  outperforms  the  FCFS scheme. This performance improvement is noticeably more significant when the number of ESUs increases to larger numbers. With the parameters used, the improvement starts when the number of ESUs is 20. This is attributed to the fact that unlike FCFS scheme, our scheme prioritizes the requests and charge the high priority requests before they expire.

\subsubsection{DCC Scheme}

Fig.~\ref{6b} gives the number of ESUs that were not able to charge before their charging request expired. It can be observed that the proposed DCC outperforms FCFS, by allowing a greater number of ESUs to charge with the available energy resources. An increase in the number of ESUs without full charge is observed when 20 ESUs compete for the available allocated community capacity. This is because DCC scheme charges the high priorty requests first. The number of ESUs leaving without full charge in FCFS is greater than that of the proposed scheme because our scheme uses priority to select the ESUs that should charge first, FCFS charges based on the time of arrival of the ESU's request. Additionally, with larger increments of the number of ESUs, the performance gap widens considerably between the two schemes.

\subsection{Communication Overhead}

We assume for the DCC that the average number of levels in the charging message by each ESU is 20, each level identifier requires 5 bytes, and the time-stamp requires 8 bytes. Our signature scheme uses elliptic curve cryptography that has smaller key sizes than the Rivest–Shamir–Adleman (RSA) scheme for the same security level. The security strength of 224-bit key in ECC is equivalent to that of 2048-bit key in the RSA cryptosystem~\cite{lenstra2000selecting}. Using an
elliptic curve additive group of order 224 bits, the signature's size is 56 bytes~\cite{raya2007securing}. Using these numbers, we calculate the size of packets in our schemes.

\subsubsection{CCC Scheme}
 The size of the one-time public key is 56 bytes, assuming that the order of $q$ is 224 bits and 56 bytes for the one-time symmetric key that is encrypted by the CC public key. In the acquisition of tokens phase in Fig.~\ref{fig:proposed}, an ESU needs to send $msg_1$ that has 56 bytes for the blinded token message, 8 bytes for the time stamp, and signature. The total size of the packet is 120 bytes. The CC should reply with $msg_2$ that contains partial blind signature that is 65 bytes. In Fig.~\ref{fig:proposed1}, a charging request $msg_3$ contains 112 bytes for the token, 56 for the BPS, 16 bytes for ciphertext of SoC and TCC, 8 bytes for the time-stamp and 56 bytes for its signature on the whole message. The total size of $msg_3$ is 248 bytes. $msg_4$ contains all charging requests within a time-slot. $msg_4$ contains $\mathcal{N}$ tokens as well as the PBS of CC signature on each token, and the cipher-text of SoC and TTC, i.e., $\mathcal{N}\times(112+16)$, also it contains $56$ for aggregated signature and 56 for the aggregator's signature on the whole message. Therefore, the total size of $msg_4$ is given as function of number of requests $\mathcal{N}$ within a time a slot as: $128 \times \mathcal{N} + 112$ bytes. Finally, $msg_5$ contains all charging schedules within a time-slot. It contains the schedule encrypted by the symmetric key of the ESU (16 bytes) and a signature on the whole message (56 bytes). Therefore, the total size of $msg_5$ as a function of $\mathcal{N}$ is given as $\mathcal{N}\times 16 + 56$ bytes.

\subsubsection{DCC Scheme}

In DCC, the size of the packet that is sent from a aggregating node to publicize the public keys is 164 bytes. The size of the homomorphic encryption ciphertext is equal to double $N$. If we choose $N$ to be 2048 bits, then the ciphertext size equals to 512 bytes. The charging report packet includes the homomorphic encryption, timestamp, and the ESU signature. The total packet size is 576 bytes.

\subsection{Computation Overhead}

 We measured the computation time of the multiplication, pairing, and exponentiation operations using Python charm cryptographic library \cite{akinyele2013charm} running on Intel Cor i5-7300HQ CPU 2.50GHz $\times$ 4 with 8 GB Ram. We used supersingular elliptic curve with the asymmetric Type 3 pairing of size 224 bits (MNT224 curve) for bilinear pairing. Our measurements indicate that the multiplication ($\textsl{Mul}$), exponentiation $(\textsl{Exp})$, and pairing $(\textsl{Pair})$ operations take 0.005, 9, and 4.4 ms, respectively. Note that the addition operation takes relatively very short time so it can be neglected. For symmetric encryption, using AES-128, the encryption operation ($\textsl{Enc}$) takes 0.0203 ms while the decryption operation ($\textsl{Dec}$) takes 0.0078 ms.

\subsubsection{CCC Scheme}  

In the acquisition of tokens phase, an ESU needs $5 \times \textsl{Mul}$= 0.025 ms. The CC needs $1 \times \textsl{Mul} + 1 \times \textsl{Pair}=4.45$ ms. To send a charging request to the aggregator, an ESU needs 
$1 \times \textsl{Mul} + 1 \times \textsl{EnC}=0.025$ ms. Upon receiving charging requsets $\mathcal{N}$, the aggreagor needs to verify a batch of signatures rather than verifying individual signatures. Batch and individual signature verifications require $\mathcal{N}+1$ and $2\mathcal{N}$ pairing operations, respectively. Finally, the CC needs to verify the aggregated signature that need $\mathcal{N}+1$ pairing operations and $\mathcal{N}$ decryption ($\textsl{Dec}$) operation. Then, preparing the charging schedules take $\mathcal{N} \times \textsl{Enc}$. Therefore, the total is $(\mathcal{N}+1)\times \textsl{Pair}+ \mathcal{N}\times\left(\textsl{Enc} + \textsl{Dnc}\right)=4.4 \times \mathcal{N}$ ms.

\subsubsection{DCC Scheme}
In DCC, the computation overhead is as follows.

\begin{itemize}
    \item ESUs: Each ESU encrypts its charging report that needs one exponential and one multiplication operations to sign its message. The ESU takes 10.15 ms to compose its charging report packet.

\item The aggregating node: verifies a batch of signatures rather than verifying individual signatures. Batch and individual signature verifications require $\mathcal{N}+1$ and $2\mathcal{N}$ pairing operations, respectively.
The aggregating node decrypts the aggregated charging messages by
computing one exponential operation. Thus, to aggregate
and sign $i$ charging report, the aggregating node needs $3\times i\times 4.4= 13.2i $ ms.
 \end{itemize}

\section{Related Work} 
\label{related}
Several works have investigated the problem of charging coordination in the smart grid, such as \cite{a5,RR61,XYZ,yang2019novel,zou2016efficient,del2016smart,6204213,ma2011decentralized,6839071,7054726}, but they do not take privacy into consideration. In \cite{a5}, a
distributed Vehicle-to-grid (V2G) control system is proposed to
satisfy the EVs charging requirements.
Tushar et. al. \cite{RR61} propose an energy management technique to encourage  EVs' owners to participate in energy trading based on a game theoretic approach.
Sortomme et. al. \cite{XYZ} proposed an algorithm to optimize energy and ancillary services scheduling. The algorithm maximizes profits to the EVs while providing additional system flexibility and peak load shaving to the utility and low costs of EV charging to the customer.

Although several privacy-preserving schemes have been proposed for the smart grid \cite{a7,e22, AMI_enahnced, 7360878}, the privacy problem we address in this paper is different.  In \cite{a7}, a privacy preserving communication protocol for power injection in smart grid have been proposed.
The aggregator collects masked bids from the storage units and sends the aggregated bid to the utility company rather than sending individual bids. The proposed scheme cannot be used for charging coordination because the CC needs the ESUs' individual charging demands.
In \cite{a8}, an authentication protocol for EVs has been proposed to protect the location of the EVs. Li et. al. \cite{lilynx} proposed an anonymous and authenticated reporting scheme for EVs.

Unlike \cite{a8} and \cite{lilynx} that address the privacy problem for EVs on roads and charging stations, we consider different network model by addressing the problem for energy units of communities. In \cite{7360878}, the authors
used a data obfuscation mechanism and proposed secure and efficient algorithms to distribute obfuscation values within an AMI network. In, a decentralized charging coordination has been proposed based on the blockchain to enable a transparent, reliable charging coordination among ESUs. However, while blockchain can reduce the reliance on intermediaries, the scheme cannot be used where there is no reliable communication, e.g., in remote areas. Also, other schemes have considered security and privacy issues

\section{Conclusion}
\label{conclusion}

In this paper, two privacy-preserving and collusion-resistant charging coordination schemes for smart grid have been proposed. Analysis, simulations, and experiments were conducted to study different attacks and evaluate the proposed schemes. The results indicate that the schemes are secure against the considered attacks, the communication and computation overheads are acceptable, and an improved performance compared with FCFS charging coordination especially at large charging demand scenario. In the CCC scheme, by acquiring anonymous and unlinkable tokens from the CC, ESUs can authenticate their charging requests anonymously and therefore a collusion attack between the CC and aggregator is mitigated. In addition, by sending multiple charging requests with random TCC and SoC that follow a truncated normal distribution, the CC can know enough data to run our charging coordination scheme, but it cannot link the data to particular ESUs. In the DCC scheme, charging is coordinated in a distributed way using data aggregation technique and homomorphic encryption to preserve ESUs' owners privacy. Moreover, more protection against collusion attacks can be achieved by increasing the number of proxy ESUs.

\begin{appendices}

\section{Sampling from Truncated Normal Distribution} \label{appendix}

The PDF of the truncated normal distribution can given by~\cite{burkardt2014truncated}.

 \begin{equation}
 \label{Eq:trunc}
\psi(\mu, s, a, b ; x)=\left\{\begin{array}{ll}{0} & {\text { if } x \leq a} \\ \frac{\phi\left(\frac{x-\mu}{s}\right)}{s\left(\Phi\left(\frac{b-\mu}{s}\right)-\Phi\left(\frac{a-\mu}{s}\right)\right)}& {\text { if } a<x<b} \\ {0} & {\text { if } b \leq x}\end{array}\right.
\end{equation}

Where  $\mu$ and ${s}$ are the mean and variance, and $a$ and $b$ specify the lower and upper truncation interval. In addition, $\phi$  and $\Phi$ are the PDF and CDF of the standard normal distribution respectively that are given as follows

\begin{equation}
\phi\left(x\right)=\frac{1}{\sqrt{2 \pi}} e^{-x^{2}/2},
\end{equation}

\begin{equation}
\begin{aligned} \Phi\left(x\right) &=\frac{1}{2}\left(1+\operatorname{erf}\left(\frac{x}{\sqrt{2}}\right)\right), \end{aligned}
\end{equation}
where erf is the standard error function~\cite{error}.

To sample from the PDF of truncated normal distribution (i.e., Eq.~\ref{Eq:trunc}), we can use inverse transform sampling method~\cite{devroye1986sample}. In this method, we first calculate the CDF of Eq.~\ref{Eq:trunc} then we get the inverse function of the CDF. Subsequently, we sample form known distribution such as uniform distribution and substituting the sample in the inverse function of the CDF. Thus, we can get a sample from the truncated normal distribution. This can be done as follows.

First, the CDF of truncated normal distribution in (\ref{Eq:trunc}) can be given by

\begin{equation}
\label{Eq: CDF truncated}
\Psi(\mu, s, a, b ; x)=\left\{\begin{array}{ll}{0} & {\text { if } x \leq a} \\ {\frac{\Phi\left(
\frac{x-\mu}{s}
\right)-\Phi\left(\frac{a-\mu}{s}\right)}{\Phi\left(\frac{b-\mu}{s}\right)-\Phi\left(\frac{a-\mu}{s}\right)}} & {\text { if } a<x<b} \\ {1} & {\text { if } b \leq x}\end{array}\right.
\end{equation}

\begin{comment}

Then, suppose we are given a value $0\leq p \leq 1$ which represents which represents the result of an evaluation of $\Psi(\overline{\mu}, \overline{s}, a, b ; x)$, and we seek $a\leq x\leq b$ satisfying 
 
$$
p=\Psi(\mu, s, a, b ; x)
$$
\end{comment}

To sample from a truncated normal distribution, we assume we have some function $rand()$ which is a source of uniform random numbers in the range [0,1] which we use to apply the inverse CDF function as follows:

$$
p=rand( )
$$
$$
x=\Psi^{-1}(\mu, s, a, b ; p)
$$

where $\Psi^{-1}$  can be calculated from Eq.~\ref{Eq: CDF truncated} to get  $x$

$$ x=\Phi^{-1}(\Phi(\frac{a-\mu}{s}) +  p * (  \Phi(\frac{b-\mu}{s})+ \Phi(\frac{a-\mu}{s})))* s+\mu$$

Here $\Phi^{-1}(y)$ is given by $\sqrt{2}$ erf $^{-1}(2 y-1)$.

\footnote{~\cite{baza11,baza1,baza6,baza3,baza7,baza2,baza5,baza8,baza12,baza13,baza10,baza4,baza14}}
\end{appendices}

\bibliographystyle{IEEEtran}
\bibliography{CC}

\end{document}